\documentclass[12pt]{article}
\pdfoutput=1

\usepackage{amsmath,amsthm,amsfonts,amssymb,srcltx,empheq}
\usepackage{graphicx}
\usepackage{here}
\usepackage{caption}
\renewcommand{\thefootnote}{\fnsymbol{footnote}}
\def\slash#1{\not\!\!#1}

\begin{document}

\title{
\begin{flushright}
\begin{minipage}{0.2\linewidth}
\normalsize
EPHOU-19-0014\\
KEK-TH-2166 \\*[50pt]
\end{minipage}
\end{flushright}
{\Large \bf 
Flavor structure of magnetized $T^2/\mathbb{Z}_2$ blow-up models
\\*[20pt]}}

\author{Tatsuo~Kobayashi$^{1}$\footnote{
E-mail address:  kobayashi@particle.sci.hokudai.ac.jp}, \ \ 
Hajime~Otsuka$^{2}$\footnote{
E-mail address: hotsuka@post.kek.jp
}, \ and \
Hikaru~Uchida$^{1}$\footnote{
E-mail address: h-uchida@particle.sci.hokudai.ac.jp}
\\*[20pt]
$^1${\it \normalsize 
Department of Physics, Hokkaido University, Sapporo 060-0810, Japan} \\
$^2${\it \normalsize 
KEK Theory Center, Institute of Particle and Nuclear Studies, KEK,}\\
{\it \normalsize 1-1 Oho, Tsukuba, Ibaraki 305-0801, Japan} \\*[50pt]}

\date{
\centerline{\small \bf Abstract}
\begin{minipage}{0.9\linewidth}
\medskip 
\medskip 
\small
We discuss the flavor structure in a resolution of magnetized $T^2/\mathbb{Z}_2$ orbifold models. 
The flavor structure of Yukawa couplings among chiral zero-modes is 
sensitive to radii of the resolved fixed points of $T^2/\mathbb{Z}_2$ orbifold. 
We show in concrete models that the realistic mass hierarchy and mixing angles of 
the quark sector can be realized at certain values of blow-up radii, 
where the number of parameters in the blow-up model can be smaller than 
toroidal orbifold one. 
\end{minipage}
}

\begin{titlepage}
\maketitle
\thispagestyle{empty}
%\clearpage
%\thispagestyle{empty}
\end{titlepage}

\renewcommand{\thefootnote}{\arabic{footnote}}
\setcounter{footnote}{0}
%\vspace{35pt}

%%%%%%%%%%%%%%%%%%%%%%%%%%%%%%%%%%%%%%%%%%%%%%%%%%%%%%%%%%%%%%%%%%%%%%%%%%%%%%%%%%%%%%%%%%%%%%%%%%%%%%%%%%%%%%%%%%%%%%%%%%%%%%%%%%%%

\section{Introduction}
Among the extra-dimensional models explaining the phenomenological and 
theoretical problems in the Standard Model (SM), toroidal orbifold models 
offer the possibility to calculate Yukawa couplings as well as matter 
wavefunctions in a controlled way. 
For example, Yukawa couplings are computed on toroidal orbifold models with magnetic fluxes \cite{Abe:2008fi,Abe:2009}\footnote{Computations on the toroidal orbifold are 
based on one on the torus with magnetic fluxes. See for Yukawa couplings on the torus with magnetic fluxes Ref.~\cite{Cremades:2004wa}.}, 
and quark and lepton masses and mixing angles as well as the CP phase are studied in Refs.~\cite{Abe:2012fj,Abe:2014,Fujimoto:2016zjs,Kobayashi:2016qag}.
In the context of string theory, such toroidal 
orbifolds~\cite{Dixon:1985jw,Dixon:1986jc} are considered the singular limit of  certain Calabi-Yau manifolds. 
To understand the nature of Calabi-Yau compactification, resolutions 
of toroidal orbifold models provide a definite way to calculate the Yukawa 
couplings analytically. 

So far, an explicit construction of metric and gauge fluxes on resolutions of toroidal orbifolds 
was performed on $\mathbb{C}^N/\mathbb{Z}_N$ with $N\geq 2$~\cite{Nibbelink:2007rd,Leung:2019oln}, where the fixed 
points are replaced by the so-called Eguchi-Hanson spaces~\cite{Eguchi:1978xp}, but the Yukawa couplings have not been derived yet. 
In Ref.~\cite{Kobayashi:2019}, the authors proposed the method of calculating 
Yukawa couplings among zero-modes on simple resolutions of $T^2/\mathbb{Z}_N$ 
orbifolds with $U(1)$ magnetic fluxes, where the orbifold fixed points 
are replaced by a part of sphere. 
The selection rules among the obtained Yukawa couplings are the same with 
the toroidal orbifold one in the blow-down limit, but Yukawa couplings are deformed  
from the toroidal orbifold results at a small but finite blow-up radius.

In this paper, we discuss phenomenological aspects of resolutions of 
$T^2/\mathbb{Z}_N$ orbifolds in a concrete three-generation model, 
where the generation numbers of elementary particles are determined by 
a magnitude of the $U(1)$ magnetic flux. 
We find that the blow-up effects sizably 
change the mass hierarchies and mixing angles of quarks to more realistic one,  compared with toroidal orbifold results. 
Furthermore, the observed values of the quark sector are well fitted by a small number of 
Higgs pairs, rather than the orbifold models. 
In this respect, blow-up models are more economical than the orbifold models.

This paper is organized as follows. 
In Section \ref{sec:2}, we briefly review the zero-mode wavefunctions on resolutions of 
$T^2/\mathbb{Z}_2$ orbifold with $U(1)$ magnetic flux, based on a six-dimensional 
gauge theory. 
The blow-up radius dependence of Yukawa couplings is discussed in Section \ref{sec:3} by using a concrete three-generation model. 
Section \ref{sec:con} is devoted to the conclusion and discussion. 
In Appendix~\ref{app}, we show the explicit calculation of the Yukawa couplings among the chiral zero-modes.

%%%%%%%%%%%%%%%%%%%%%%%%%%%%%%%%%%%%%%%%%%%%%%%%%%%%%%%%%%%%%%%%%%%%%%%%%%%%%%%%%%%%%%%%%%%%%%%%%%%%%%%%%%%%%%%%%%%%%%%%%%%%%%%%%%%%

\section{Zero-mode wavefunctions on resolutions of $T^2/\mathbb{Z}_2$ orbifold}
\label{sec:2}
In this section, we briefly review the chiral zero-mode wavefunctions, starting from the  six-dimensional gauge theory on toroidal background including its resolutions with constant magnetic fluxes.

%%%%%%%%%%%%%%%%%%%%%%%%%%%%%%%%%%%%%%%%%%%%%%%%%%%%%%%%%%%%%%%%%%

\subsection{Wavefunctions on $T^2$ and $T^2/\mathbb{Z}_N$}
\label{subsec:2_1}
First of all, we briefly review the chiral zero-mode wavefunctions on the two-dimensional torus $T^2$ with $U(1)$ magnetic flux~\cite{Cremades:2004wa}, where $T^2 \simeq \mathbb{C}/\Lambda$, $\Lambda$ is a two-dimensional lattice spanned by $\alpha_1=2\pi R$ and $\alpha_2=2\pi R \tau$, $R \in \mathbb{R}$ is the radius and $\tau \in \mathbb{C}$ is  a complex structure of the torus.
The metric on $T^2$ in the complex coordinates $z=x+\tau y$ is given by
\begin{align}
g_{i\bar{j}}=
(2\pi R)^2 
\begin{pmatrix}
0 & \frac{1}{2} \\
\frac{1}{2} & 0 \\
\end{pmatrix}
,
\end{align}
where $i=z, \bar{z}$.
The $U(1)$ magnetic flux on $T^2$ is given by
\begin{align}
F=\frac{i\pi M}{{\rm Im}\tau} dz\wedge d\bar{z}, 
\end{align}
which is quantized on $T^2$, namely $(2\pi)^{-1}\int_{T^2}F=M\in \mathbb{Z}$. 
In our analysis, we employ the following background gauge field
\begin{align}
A=\frac{\pi M}{{\rm Im}\tau} {\rm Im}(\bar{z}dz), 
\end{align}
leading to the quantized flux. 
Hereafter, we focus on the case without Wilson-lines for simplicity.
On this gauge background, the zero-mode wavefunction of fermion with the unit charge, $q=1$,
\begin{align}
\Psi(z,\bar{z}) = 
\begin{pmatrix}
\psi_+\\
\psi_-\\
\end{pmatrix}
\end{align}
is a solution of the following zero-mode Dirac equation
\begin{align}
\slash{D}\Psi=0. 
\end{align}
Note that either component $\psi_+$ or $\psi_-$ has a solution of the above zero-mode equation since the magnetic flux $|M|$ generates the net-number of chirality.
In particular, when $M$ is positive (negative), 
$\psi_+$ ($\psi_-$) has $|M|$ number of degenerate zero-modes described by
\begin{align}
\psi_+^{j,M}(z)&={\cal N}^{j,M} e^{i\pi Mz{\rm Im}(z)/{\rm Im}\tau} 
\vartheta
\begin{bmatrix}
\frac{j}{M}\\
0
\end{bmatrix}
(Mz, M\tau) \qquad (M>0),
\nonumber\\
\psi_-^{j,|M|}(z)&={\cal N}^{j,|M|} e^{i\pi |M|z{\rm Im}(z)/{\rm Im}\tau} 
\vartheta
\begin{bmatrix}
\frac{j}{|M|}\\
0
\end{bmatrix}
(|M|z, |M|\tau) \qquad (M<0),
\label{eq:T2wavefcn}
\end{align}
with $j=0,1,...,(|M|-1)$, where $\vartheta$ denotes the Jacobi theta function
\begin{align}
\vartheta
\begin{bmatrix}
a\\
b
\end{bmatrix}
(z, \tau)
=
\sum_{l\in \mathbb{Z}}
e^{\pi i (a+l)^2\tau}
e^{2\pi i (a+l)(z+b)}
,
\end{align}
and ${\cal N}^{j,|M|}$ denotes the normalization factor
\begin{align}
{\cal N}^{j,|M|}=\left(\frac{2|M|{\rm Im}\tau}{{\cal A}^2}\right)^{1/4}
\end{align}
with the area of the torus ${\cal A}=4\pi^2 R^2 {\rm Im}\tau$.
Hereafter, we set $2\pi R=1$. 
The zero-mode wavefunction of scalar field $\phi^{j,|M|}(z)$
%, derived from the internal component of the 6-dimensional vector field,
is also the same functional form with massless fermion. 

Next, we move on to the chiral zero-mode wavefunctions on the toroidal orbifold $T^2/\mathbb{Z}_2$, constructed by further identifying $T^2$ with $\mathbb{Z}_2$-transformation $z\rightarrow -z$. Under the transformation, there exist four-fixed points corresponding to the orbifold singularities.
The $\mathbb{Z}_2$-even and -odd zero-mode wavefunctions are described by~\cite{Abe:2008fi}
\begin{align}
\psi_{T^2/\mathbb{Z}_2^\pm}^{j,|M|} = 
\left\{
\begin{array}{r}
{\cal N}^{j,|M|} e^{i\pi |M| z {\rm Im}(z)/{\rm Im}(\tau)} 
\vartheta
\begin{bmatrix}
\frac{j}{|M|}\\
0
\end{bmatrix}
(|M|z, |M|\tau) \qquad (j=0,\frac{|M|}{2})
\\
\frac{{\cal N}^{j,|M|}}{\sqrt{2}} e^{i\pi |M| z {\rm Im}(z)/{\rm Im}(\tau)}
\left(
\vartheta
\begin{bmatrix}
\frac{j}{|M|}\\
0
\end{bmatrix}
(|M|z, |M|\tau) \pm 
\vartheta
\begin{bmatrix}
\frac{|M|-j}{|M|}\\
0
\end{bmatrix}
(|M|z, |M|\tau)
\right)\\
(0<j<\frac{|M|}{2})
\end{array}
\right.
,
\label{eq:T2Z2}
\end{align}
where the Jacobi theta function $\vartheta^{j,|M|}(z)$ satisfies $\vartheta^{j,|M|}(-z)=\vartheta^{|M|-j, |M|}(z)$ and $\vartheta^{j,|M|}(z)$ with $j=0, |M|/2$ show $\mathbb{Z}_2$-even modes.
In the case that $M$ is even, the total numbers of $\mathbb{Z}_2$-even and -odd zero-modes are $(|M|/2+1)$ and $(|M|/2-1)$, respectively. In the case that $M$ is odd, the total numbers of $\mathbb{Z}_2$-even and -odd zero-modes are $((|M| - 1)/2 + 1)$ and $((|M| - 1)/2)$, respectively. Moreover, the $\mathbb{Z}_2$ projection makes either $\mathbb{Z}_2$ -even or -odd modes remain. 
We show the numbers of the zero-modes for each magnetic flux $M$ explicitly in TABLE \ref{tb:zeromode}, 
\begin{table}[h]
\centering
    \begin{tabular}{|c|ccccccccccccc|} \hline 
    $|M|$ &\ 0\ &\ 1\ &\ 2\ &\ 3\ &\ 4\ &\ 5\ &\ 6\ &\ 7\ &\ 8\ &\ 9\ & 10 & 11 & 12\ \\ \hline
	even & 1 & 1 & 2 & 2 & 3 & 3 & 4 & 4 & 5 & 5 & 6 & 6 & 7 \\
	odd & 0 & 0 & 0 & 1 & 1 & 2 & 2 & 3 & 3 & 4 & 4 & 5 & 5 \\ \hline
	\end{tabular}
	\caption{
		The total numbers of degenerate zero-modes for $\mathbb{Z}_2$ -even and -odd wavefunctions.}
	\label{tb:zeromode}
\end{table}
from which it turns out that several magnetic fluxes give three degenerate zero-modes 
identified with the three generations of quarks and leptons. 
Note that in the case without orbifolding, three degenerate zero-modes appear only for the magnetic flux $|M| = 3$. 

%%%%%%%%%%%%%%%%%%%%%%%%%%%%%%%%%%%%%%%%%%%%%%%%%%%%%%%%%%%%%%%%%%

\subsection{Blow-up of $T^2/\mathbb{Z}_2$}
\label{subsec:2_2}
\begin{figure}[H]
\centering
  \begin{minipage}{0.4\columnwidth}
  \centering
     \includegraphics[scale=0.2]{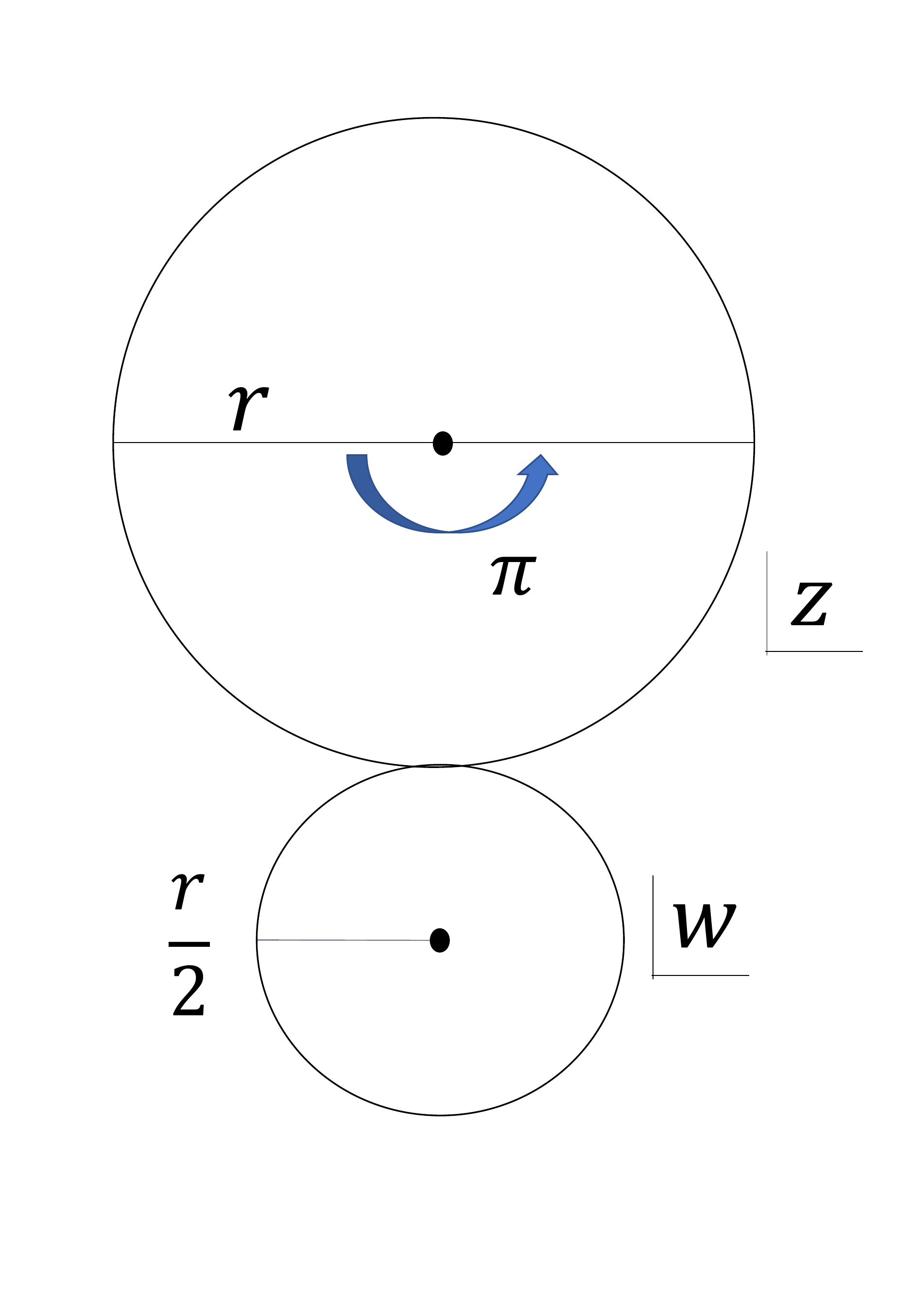}
  \end{minipage}
  \begin{minipage}{0.4\columnwidth}
   \centering
    \includegraphics[scale=0.3]{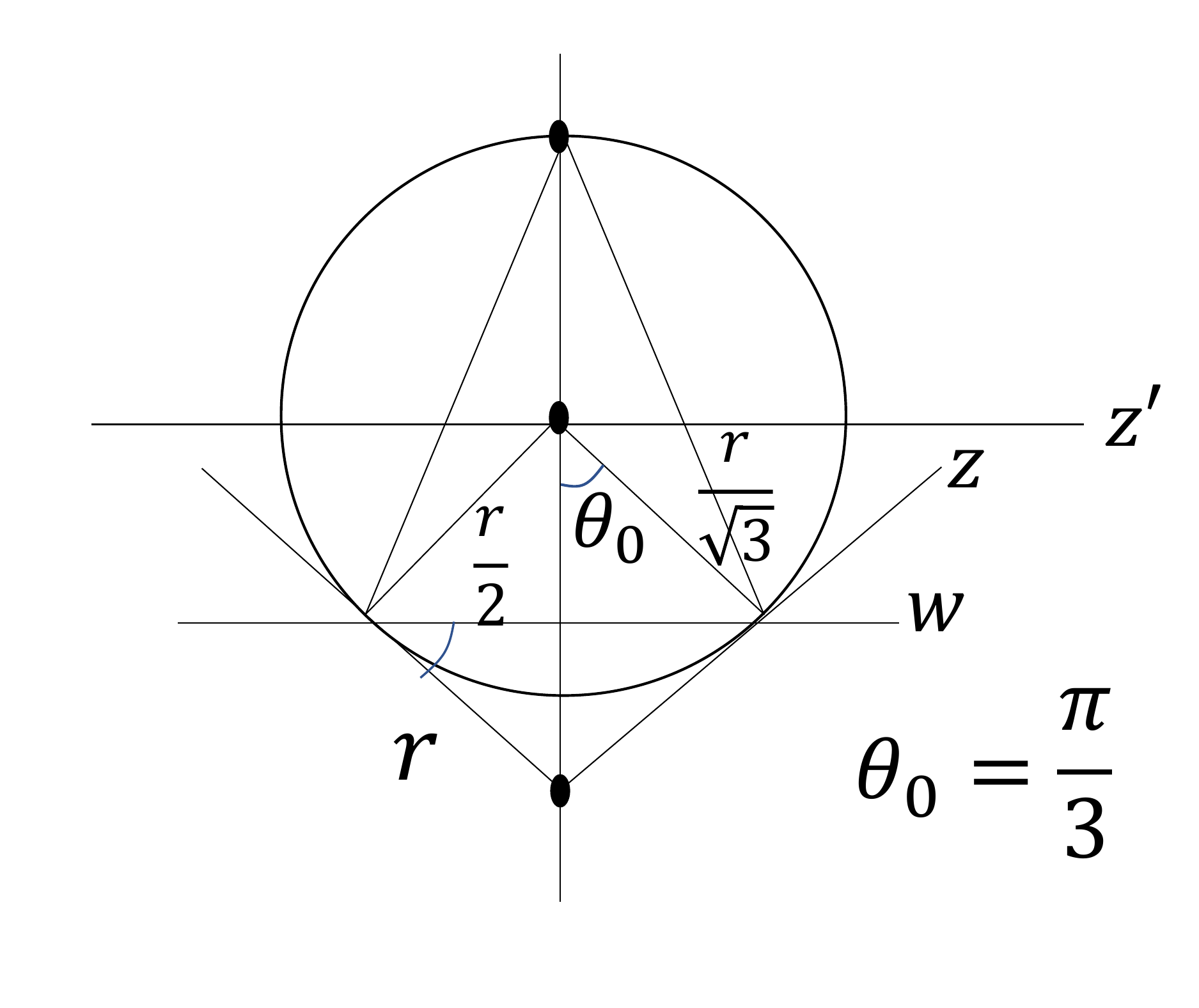}
  \end{minipage}
    \caption{The left panel shows the net of a cone cut out from $T^2/\mathbb{Z}_2$ with the coordinates of $T^2/\mathbb{Z}_2$ $z$ and the cut edge $w$, whereas in the right panel, the cone cut out by $T^2/\mathbb{Z}_2$ whose coordinate is $z$ and a section of embedded quarter of $S^2$ with radius $r/\sqrt{3}$ whose coordinate is $z'$ or $w$ are drawn.}
    \label{fig:Z2}
\end{figure}

Here, we review the zero-mode wavefunctions on the resolutions of $T^2/\mathbb{Z}_2$ orbifold with $U(1)$ magnetic flux, where the orbifold fixed points are replaced by a part of two-dimensional sphere $S^2$~\cite{Kobayashi:2019}\footnote{It is the reason why we use $S^2$ that the Euler number of $T^2/\mathbb{Z}_2$ is same as one of $S^2$.}, as shown in FIG. \ref{fig:Z2}.
In particular, we cut a cone with the generatrix length $r$ 
whose apex is a fixed point and embed a fourth of $S^2$ with radius $r/\sqrt{3}$ there. Note that the radius of the cut edge is $r/2$ due to the $\mathbb{Z}_2$ identification $z\simeq -z$ \footnote{We also note that the deficit angles of all the fixed points are the same, indicating that a fourth of $S^2$ with radius $r/\sqrt{3}$ is embedded into each fixed point. Hence, at the end of the day, all the $S^2$ region is pasted together, which means that the resolutions of $T^2/\mathbb{Z}_2$ orbifold is homeomorphic to $S^2$.}. 
In what follows, we focus on the fixed point $z_I=0$ on $T^2/\mathbb{Z}_2$ orbifold.

Let us move on to the zero-mode wavefunctions on the resolution of $T^2/\mathbb{Z}_2$ orbifold through the above geometrical procedure.
As discussed in Ref.~\cite{Conlon:2008qi}, the zero-mode wavefunction of scalar field on $\mathbb{CP}^1\simeq S^2$ with $U(1)$ magnetic flux $M'$ is described by
\begin{align}
\phi^{j, |M'|}_{S^2} (z') = {\cal N'}^{j,|M'|} \frac{f^{j,|M'|}(z')}{\left(1+|z'|^2\right)^{|M'|/2}},
\end{align}
where $f^{j, |M'|}(z')$ is the holomorphic function in the coordinate of $\mathbb{CP}^1$ $z'$ with the zero-mode index $j$ determined by the flux $|M'|$ and ${\cal N'}^{j,|M'|}$ is the normalization factor. 
The reason why we take the same index $j$ appearing in $T^2/\mathbb{Z}_2$ wavefunctions 
is to smoothly connect $S^2$ and $T^2/\mathbb{Z}_2$ wavefunctions in the blow-down limit $r\rightarrow 0$. 
The zero-mode wavefunction of fermion is also the same functional form but the meaning of $M'$ is modified to the effective magnetic flux due to the curvature of $S^2$.
The $S^2$ coordinate $z'$ is related to the $T^2/\mathbb{Z}_2$ coordinate $z$ through another coordinate of $S^2$ $w$, which is written by  $w=\frac{r}{2}\sqrt{3}z'=\frac{r}{2}e^{i\varphi}$ 
at $\theta =\theta_0$ while $z=re^{i\varphi/2}$ at the same point as shown in FIG.~\ref{fig:Z2}, where $z'=\tan\frac{\theta}{2}e^{i\varphi}$ in the setup of FIG.~\ref{fig:Z2}.

Thus, in order for the $T^2/\mathbb{Z}_2$ wavefunction at $z=re^{i\varphi/2}$ smoothly to connect with the $S^2$ wavefunction at $z'=\frac{1}{\sqrt{3}}e^{i\varphi}$, the following conditions must be satisfied
\begin{eqnarray}
\phi_{T^2/\mathbb{Z}_2^{\pm}}^{j, |M|}(z)\biggl|_{z=r e^{i\varphi/2}} &=& 
\phi^{j, |M'|}_{S^2} (z')\biggl|_{z'=\frac{1}{\sqrt{3}} e^{i\varphi}},
\nonumber\\
\frac{d \phi_{T^2/\mathbb{Z}_2^{\pm}}^{j, |M|}(z)}{d z}\biggl|_{z=r e^{i\varphi/2}} &=& \frac{1}{\frac{\sqrt{3}}{2}r} \frac{d \phi^{j, |M'|}_{S^2} (z')}{d z'}\biggl|_{z'=\frac{1}{\sqrt{3}} e^{i\varphi}}. 
\label{eq:concon}
\end{eqnarray}
Here, we take the Ansatz in Ref.~\cite{Kobayashi:2019} that the holomorphic part of $T^2/\mathbb{Z}_2$ wavefunction (\ref{eq:T2Z2}), denoted by the following $h^{j,|M|}(z)$
\begin{align}
h^{j, |M|} (z) = 
\left\{
\begin{array}{c}
g^{j, |M|} (z), \qquad (j = 0, |M|/2,\ \mathbb{Z}_2 \text{-even})
\\
\frac{g^{j, |M|} (z) + g^{j, |M|-j} (z)}{\sqrt{2}}, \qquad (0 < j < |M|/2,\ \mathbb{Z}_2 \text{-even})
\\
\frac{g^{j, |M|} (z) + g^{j, |M|-j} (z)}{\sqrt{2}}, \qquad (0 < j < |M|/2,\ \mathbb{Z}_2 \text{-odd})
\end{array}
\right.
\label{eq:holh}
\end{align}
with
\begin{align}
g^{j, |M|}(z) \equiv e^{\frac{\pi |M|}{2{\rm Im}\tau}z^2}
\vartheta
\begin{bmatrix}
\frac{j}{|M|}\\
0
\end{bmatrix}
(|M|z, |M|\tau),
\label{eq:holg}
\end{align}
is unchanged after the blow-up. 
Under the coordinate transformation $z\rightarrow w$, the transformation of the derivative of $S^2$ holomorphic function $f^{j, |M'|}$ is provided by
\begin{align}
\frac{d f^{j, |M'|}(z)}{d z}\biggl|_{z=r e^{i\varphi/2}} = \frac{d f^{j, |M'|}(w)}{d w}\biggl|_{z=\frac{r}{2} e^{i\varphi}} = 
\frac{1}{\frac{\sqrt{3}}{2}r} \frac{d f^{j, |M'|}(\frac{\sqrt{3}}{2}r z')}{d z'}\biggl|_{z'=\frac{1}{\sqrt{3}} e^{i\varphi}}. 
\label{eq:derivf}
\end{align}
Then, from the non-holomorphic parts in the condition for the continuous connection  (\ref{eq:concon}), the following conditions are obtained
\begin{align}
\frac{{\cal N}^{j,|M|}_1}{{\cal N'}^{j,|M'|}_1} = \left(\frac{\sqrt{3}}{2}\right)^{|M'|}e^{\frac{|M'|}{4}},\qquad 
\frac{|M'|}{4}=\frac{\pi r^2}{2{\rm Im}\tau}|M|.
\label{eq:ratio_N}
\end{align}
The former condition shows the relation between the normalization factor of the $T^2/\mathbb{Z}_2$ part ${\cal N}^{j,|M|}_1$ and that of the $S^2$ part ${\cal N'}^{j,|M'|}_1$, and the latter condition shows that the embedded effective flux equals to the cut-out effective flux \footnote{It is because that the area cut out from $T^2/\mathbb{Z}_2$ with the effective flux $M$ is $\pi r^2/2$ compared with the total area ${\rm Im}\tau$ and the a fourth of $S^2$ with $M'$ is embedded.}. 
In other words, the amount of the effective flux quanta is unchanged through the blow-up.
On the other hand, from the Ansatz and the coordinate transformation (\ref{eq:derivf}), the $S^2$ holomorphic function $f^{j,|M'|}(z')$ is determined by using Eq. (\ref{eq:holg}) as the following
\begin{align}
f^{j, |M'|} (z') = 
\left\{
\begin{array}{c}
g^{j, |M|}\left(\frac{\sqrt{3}}{2}r z'\right), \qquad (j = 0, |M|/2,\ \mathbb{Z}_2 \text{-even})
\\
\sqrt{2} g^{j, |M|}\left(\frac{\sqrt{3}}{2}r z'\right), \qquad (0 < j < |M|/2,\ \mathbb{Z}_2 \text{-even})
\\
0, \qquad (0 < j < |M|/2,\ \mathbb{Z}_2 \text{-odd})
\end{array}
\right.
,
\label{eq:holf}
\end{align}
where we use the facts that $\vartheta^{|M|-j,M}(z)=\vartheta^{j, |M|}(ze^{i \pi})$ and the argument of $z'$ is twice as much as that of $z$.
Recall that $M'$ is related to $M$ through Eq. (\ref{eq:ratio_N}).
The above holomorphic functions show that only $\mathbb{Z}_2$-even mode is uplifted to $S^2$ and they are consistent with the fact that the $\mathbb{Z}_2$-odd mode vanishes at the fixed point $z_I=0$ in the blow-down limit $r \rightarrow 0$.
Note that the same applies to the case of the blow-up of the singularities at $z_I=1/2,\tau/2$ because of $g^{j,|M|}(-z_I)=g^{j,|M|}(z_I)$, while in the case of the blow-up of the singularity at $z_I=(\tau+1)/2$, only $\mathbb{Z}_2$-even (-odd) mode is uplifted to $S^2$ when the flux $M$ is even (odd) because of $g^{j,|M|}(-z_I)=e^{i \pi |M|}g^{j,|M|}(z_I)$.

As a result, the zero-mode wavefunctions after the blow-up of the singularity at $z_I=0$ are described by
\begin{align}
\phi^{j,|M|}_{\rm up} = 
\left\{
\begin{array}{c}
\phi^{j,|M'|}_{S^2} = 
\cfrac{{\cal N'}^{j,|M'|}_1}{\left(1+|z'|^2\right)^{\frac{|M'|}{2}}}f^{j,|M'|}(z')\qquad (|z'|\leq \frac{1}{\sqrt{3}})
\\  
\phi^{j, |M|}_{T^2/\mathbb{Z}_2}= {\cal N}^{j,|M|}_1 e^{-\frac{\pi |M| |z|^2}{2{\rm Im}\tau}} h^{j, |M|}(z)\qquad (|z|\geq r)
\end{array}
\right.
,
\label{eq:upwave}
\end{align}
with Eq. (\ref{eq:ratio_N}), where the $S^2$ holomorphic function $f^{j,|M'|}(z')$ and the $T^2/\mathbb{Z}_2$ holomorphic function $h^{j, |M|}(z)$ are shown in Eq. (\ref{eq:holf}) and Eq. (\ref{eq:holh}), respectively.
Note that the fermionic zero-mode wavefunctions $\psi^{j,|M|}_{\rm up}$ are also the same functional forms, but the meaning of $M'$ is modified to the effective magnetic flux.
The normalization factors are calculated by the wavefunctions (\ref{eq:upwave}),
\begin{align}
\left|{\cal N}^{j,|M|}_1\right| \simeq
\left\{
\begin{array}{c}
\left|{\cal N}^{j,|M|}\right|, \qquad (j = 0, |M|/2,\ \mathbb{Z}_2 \text{-even})
\\
\left|{\cal N}^{j,|M|}\right| \left(1 - \frac{1}{2\pi} \left(\frac{\pi r^2}{2}\right)^2 \left|\left[\phi^{j,|M|}_{T^2/\mathbb{Z}_2^-}\right]' \left(0\right)\right|^2 \right), \qquad (0 < j < |M|/2,\ \mathbb{Z}_2 \text{-even}) \\
\left|{\cal N}^{j,|M|}\right| \left(1 + \frac{1}{2\pi} \left(\frac{\pi r^2}{2}\right)^2 \left|\left[\phi^{j,|M|}_{T^2/\mathbb{Z}_2^-}\right]' \left(0\right)\right|^2 \right), \qquad (0 < j < |M|/2,\ \mathbb{Z}_2 \text{-odd})
\end{array}
\right.
,
\label{eq:Normalizationapp}
\end{align}
where ${\cal N}^{j,|M|}=(2|M|{\rm Im}(\tau)/{\cal A}^2)^{1/4}$ is the normalization factor on $T^2/\mathbb{Z}_2$ before the blow-up and $\phi^{j,|M|}_{T^2/\mathbb{Z}_2^-}$ is the $\mathbb{Z}_2$-odd wavefunction on $T^2/\mathbb{Z}_2$ before the blow-up.
Note that
$\left[\phi^{j,|M|}_{T^2/\mathbb{Z}_2^+}\right]' \left(0\right) \equiv \frac{d\phi^{j,|M|}_{T^2/\mathbb{Z}_2^+}}{dz} \Biggl|_{z=0} = 0$.
Furthermore, the normalization factor of the $S^2$ part is obtained from the normalization relation (\ref{eq:ratio_N}). 
Note that the blow-up radius should be smaller than $1/2$ to ensure our analysis, 
otherwise the blow-up region exceeds the bulk region

So far, we have focused on $U(1)$ magnetic flux background, but it is easy to extend the case with the following $U(N)$ magnetic flux,
\begin{align}
F_{z\bar{z}}=\frac{i\pi}{{\rm Im}\tau}
\begin{pmatrix}
M_a \mathbb{I}_{N_a} & \ \\
\ & M_b \mathbb{I}_{N_b}
\end{pmatrix}, 
\end{align}
where $N_a + N_b = N$ and $\mathbb{I}_{N_{a,b}}$ denotes the $(N_{a,b} \times N_{a,b})$ identity matrix, which breaks the $U(N)$ gauge symmetry to $ U(N_a) \times U(N_b)$.
Under this gauge background, the chiral zero-mode wavefunctions with the representation $(N_a,\bar{N_b})$ are given by the same functional forms with the $U(1)$ gauge background but the flux $M$ is written by $M = M_a - M_b$.

%%%%%%%%%%%%%%%%%%%%%%%%%%%%%%%%%%%%%%%%%%%%%%%%%%%%%%%%%%%%%%%%%%%%%%%%%%%%%%%%%%%%%%%%%%%%%%%%%%%%%%%%%%%%%%%%%%%%%%%%%%%%%%%%%%%%

\clearpage

\section{The model}
\label{sec:3}
In this section, we study a model containing three generations of quarks on the resolutions of the toroidal orbifold background, where the orbifold fixed points are blown up. 

As an illustrating model, we use the flavor model in Ref.~\cite{Abe:2014}.
This model is ten-dimensional $U(8)$ supersymmetric Yang-Mills theory 
compactified on $T^6/(\mathbb{Z}_2 \times \mathbb{Z}_2)$.
The six-dimensional $T^6$ is written by a product of three $T^2$'s, i.e. $T^6 = T^2_1 \times T^2_2 \times T^2_3$, where we denote the 
$i$-th complex coordinate on $T^2_i$ by $z_i$.
There are three adjoint matter fields $\Phi_i$ ($i=1,2,3$) in the terminology of four-dimensional supersymmetric theory, where 
$\Phi_i$ have the vector index on $z_i$.
These chiral matter fields correspond to the left-handed and right-handed 
fermions and Higgs fields.
For example, the left-handed (right-handed) quarks and leptons are originated from $\Phi_1$ ($\Phi_2$), while the up and down Higgs fields correspond to 
$\Phi_3$.
In the model of Ref.~\cite{Abe:2014}, magnetic fluxes of $U(8)$ are introduced in the extra dimensions such that three generations of 
quarks and leptons are realized and four-dimensional ${\cal N}=1$ supersymmetry remains.
Zero-mode wavefunctions of each $\Phi_i$ are written by 
products of three wavefunctions on $z_j$, $\Phi_i=\prod_{j=1}^3\phi_{i(j)}(z_j)$.
The left-handed and right-handed quarks and leptons have three zero-modes 
on one of $z_j$, say $z_1$, while their zero-modes are single functions 
on the other four-dimensional space.
That is, the flavor structure is determined by wavefunctions on $T^2_1$, while the other parts are relevant to only the overall factors of 
Yukawa couplings.
Thus, here we concentrate on only the first two-dimensional part, $T^2_1/\mathbb{Z}_2$.
Then, we use the six-dimensional model on $T^2/\mathbb{Z}_2$, which leads to the same mass ratios and mixing angles as the model 
in Ref.~\cite{Abe:2014} in order to illustrate blow-up effects 
on mass matrices.\footnote{Study on blow-up of $T^6/(\mathbb{Z}_2 \times \mathbb{Z}_2)$ is beyond our scope.}
In particular, we focus on the quark sector\footnote{Of course, we can 
discuss the Yukawa couplings of the lepton sector, but it highly depends on the structure of 
neutrino sector such as Majorana mass etc. 
See e.g. Ref.~\cite{Kobayashi:2015siy}.
We thus postpone the detailed analysis of lepton sector for a future work.}.

In the six-dimensional supersymmetric theory, 
$\Phi_1$ corresponds to the extra dimensional direction of the vector 
multiplet, while $\Phi_{2,3}$ are hypermultiplets.
%\footnote{
%However, our purpose is to examine the six-dimensional $T^2/\mathbb{Z}_2$ model, which 
%leads to the same quark mass matrices as \cite{Abe:2014}.
%We do not insist on supersymmetry, but we focus just on fermionic fields in
%$\Phi_1$ and $\Phi_2$ as left-handed and right-handed quarks and 
%scalar fields in $\Phi_3$ as the up and down Higgs fields.}
We set the magnetic fluxes and $\mathbb{Z}_2$ parity such that 
the magnetic flux $M=-5$ ($-7$) appears in the zero-mode equation for 
the left-handed (right-handed) quarks and they are $\mathbb{Z}_2$ even (odd), 
while the magnetic flux $M=12$ appears in the zero-mode equation 
for the up and down Higgs fields.
Then,  it turns out that quarks have three degenerate zero-modes, whereas there exist five degenerate zero-modes for the up and down Higgs sectors. 
TABLE \ref{tb:model} shows these behavior.

\begin{table}[h]
\centering
    \begin{tabular}{|c|c|c|c|} \hline 
    \par  &  Left-handed quarks & Right-handed quarks & Higgs \\ \hline
	$M$ ~($\mathbb{Z}_2$ parity) &  $-5$\ ~(even) & $-7$\ ~(odd) & 12\ ~(odd) \\ \hline
%	Lepton & -4\ (even) & -8\ (odd) & 12\ (odd) \\ \hline
	\end{tabular}
	\caption{
		Magnetic fluxes for left- and right-handed quarks and Higgs  sectors.}
	\label{tb:model}
\end{table}
Each wavefunction is described as in Eq. (\ref{eq:upwave}), where the label $j$ is replaced by $I\ (0 \leq I \leq 2)$, $J+1\ (0 \leq J \leq 2)$, and $K+1\ (0 \leq K \leq 4)$ for the left-handed quarks, the right-handed quarks, and the Higgs fields, respectively. 
Then, Yukawa couplings of the quark sector are written by
\begin{align}
Y_{IJK}H_{K}Q_{LI}Q_{RJ}=
(Y_{IJ0}H_{0}+Y_{IJ1}H_{1}+Y_{IJ2}H_{2}+Y_{IJ3}H_{3}+Y_{IJ4}H_{4})Q_{LI}Q_{RJ},
\end{align}
where $Q_{LI}$, $Q_{RJ}$, and $H_{K}$ are the four-dimensional left-handed and right-handed quarks and the Higgs fields, respectively and $Y_{IJK}$ is calculated as follows:
\begin{eqnarray}
Y_{IJK}
&=&
\int_{T^2/\mathbb{Z}_2} dzd\bar{z}\ \phi^{I,5}_{T^2/\mathbb{Z}_2^+}(z) \phi^{J+1,7}_{T^2/\mathbb{Z}_2^-}(z) \left(\phi^{K+1,12}_{T^2/\mathbb{Z}_2^-}(z)\right)^{\ast} \notag \\
&-&
\sum_{z_I}
\int_{\left|z-z_I\right| \leq r} dzd\bar{z}\ \phi^{I,5}_{T^2/\mathbb{Z}_2^+}(z) \phi^{J+1,7}_{T^2/\mathbb{Z}_2^-}(z) \left(\phi^{K+1,12}_{T^2/\mathbb{Z}_2^-}(z)\right)^{\ast},
\label{eq:Yukawa}
\end{eqnarray}
where the first term is the whole $T^2/\mathbb{Z}_2$ term and the second term is the cut-out term from $T^2/\mathbb{Z}_2$.
In addition, the terms on the $S^2$ regions after the blow-up of the singularities are needed originally but all of them become zero because the wavefunctions of the right-handed quarks and Higgs fields vanish on the $S^2$ regions after the blow-up of the singularities at $z_I = 0, 1/2, \tau/2$ and the wavefunctions of the left-handed quarks and Higgs fields vanish on the $S^2$ region after the blow-up of the singularities at $z_I = (\tau+1)/2$.
The explicit form of the first term in Eq. (\ref{eq:Yukawa}) is shown in Ref.~\cite{Abe:2009,Abe:2014}, and the second term is calculated
\begin{eqnarray}
&&\int_{\left|z-z_i\right| \leq r} dzd\bar{z}\ \phi^{I,5}_{T^2/\mathbb{Z}_2^+}(z) \phi^{J+1,7}_{T^2/\mathbb{Z}_2^-}(z) \left(\phi^{K+1,12}_{T^2/\mathbb{Z}_2^-}(z)\right)^{\ast} \notag \\
&\simeq& \frac{1}{2\pi} \left(\frac{\pi r^2}{2}\right)^2 \left(\phi^{I,5}_{T^2/\mathbb{Z}_2^+} \phi^{J+1,7}_{T^2/\mathbb{Z}_2^-}\right)' (z_I) \left(\phi'^{k,M}_{T^2/\mathbb{Z}_2^-}(z_I)\right)^{\ast},
\label{eq:second}
\end{eqnarray}
whose explicit form is shown in Appendix~\ref{app}. 
In the following analysis, we focus on two cases: blowing up all the fixed points, each with the same blow-up radius in Section~\ref{sec:3_1} and  different blow-up radii in Section~\ref{sec:3_2}.

%%%%%%%%%%%%%%%%%%%%%%%%%%%%%%%%%%%%%%%%%%%%%%%%%%%%%%%%%%%%%%%%%%%%%%%%%%%%%%%%%%%%%%%%%%%%%%%%%%%%%%%%%%%%%%%%%%%%%%%%%%%%%%%%%%%%
\subsection{Model 1}
\label{sec:3_1}
\begin{figure}[H]
\centering
  \begin{minipage}{0.41\columnwidth}
  \centering
     \includegraphics[scale=0.35]{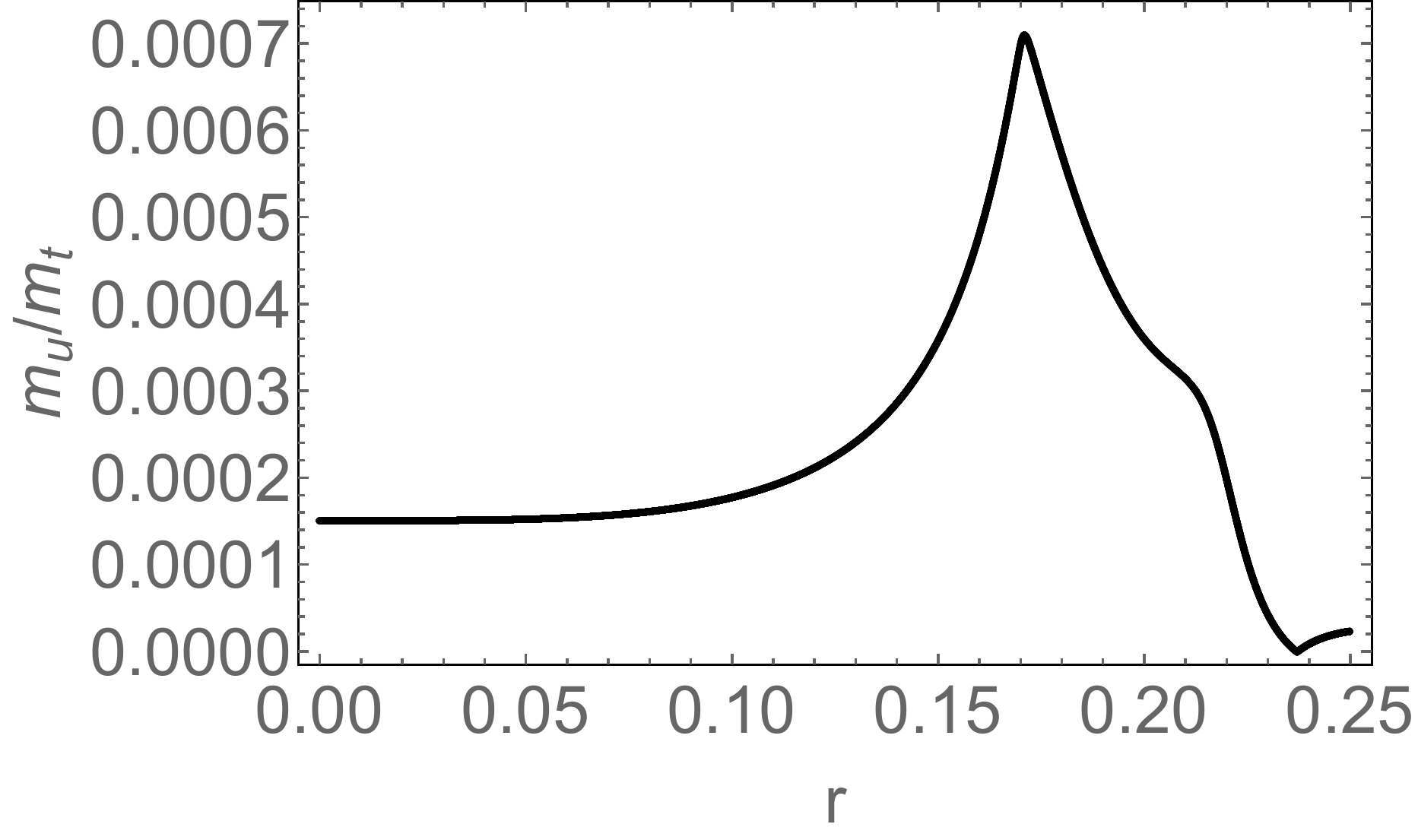}
  \end{minipage}
  \begin{minipage}{0.4\columnwidth}
  \centering
     \includegraphics[scale=0.35]{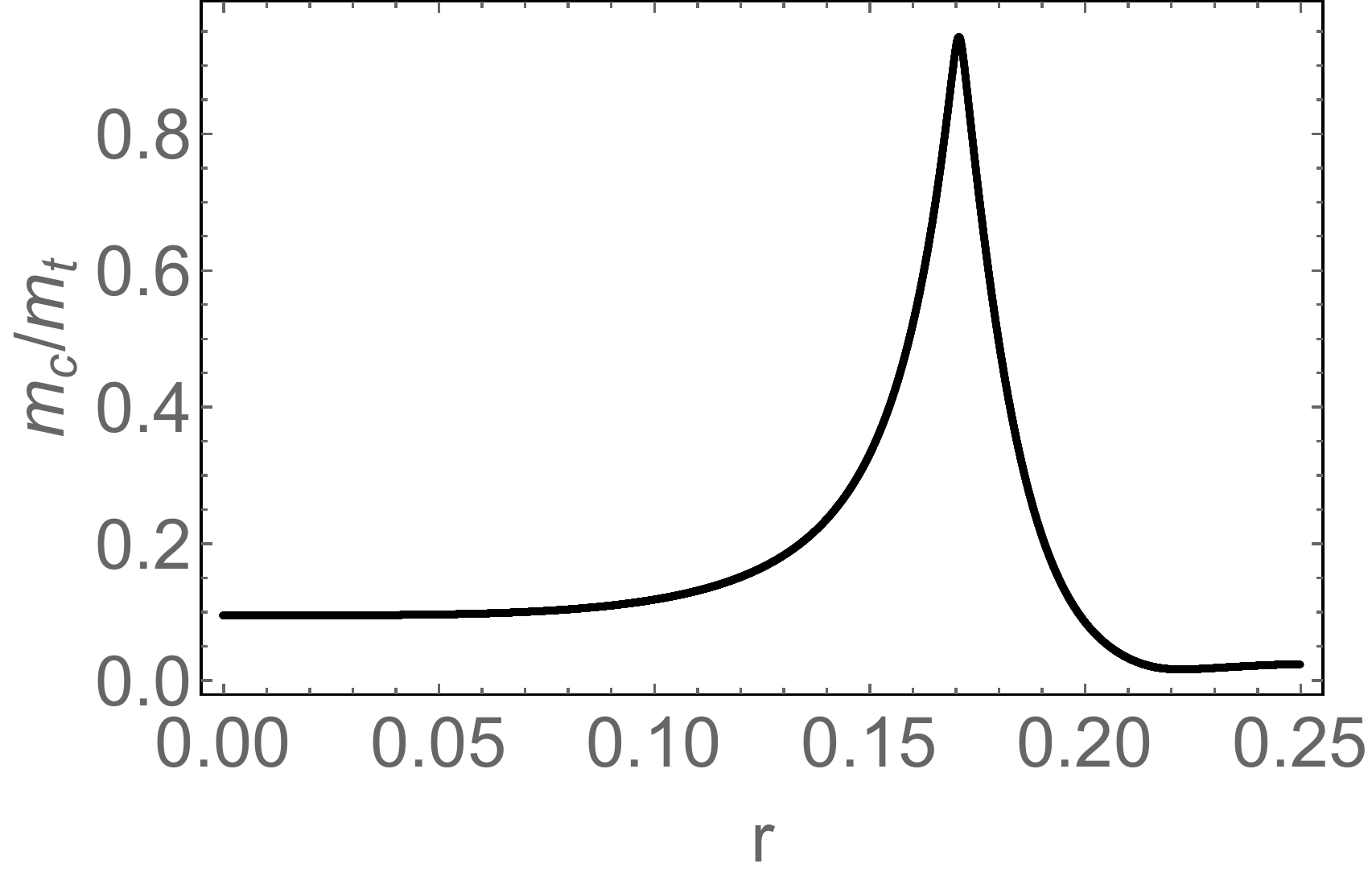}
  \end{minipage} \\
  \begin{minipage}{0.41\columnwidth}
  \centering
     \includegraphics[scale=0.35]{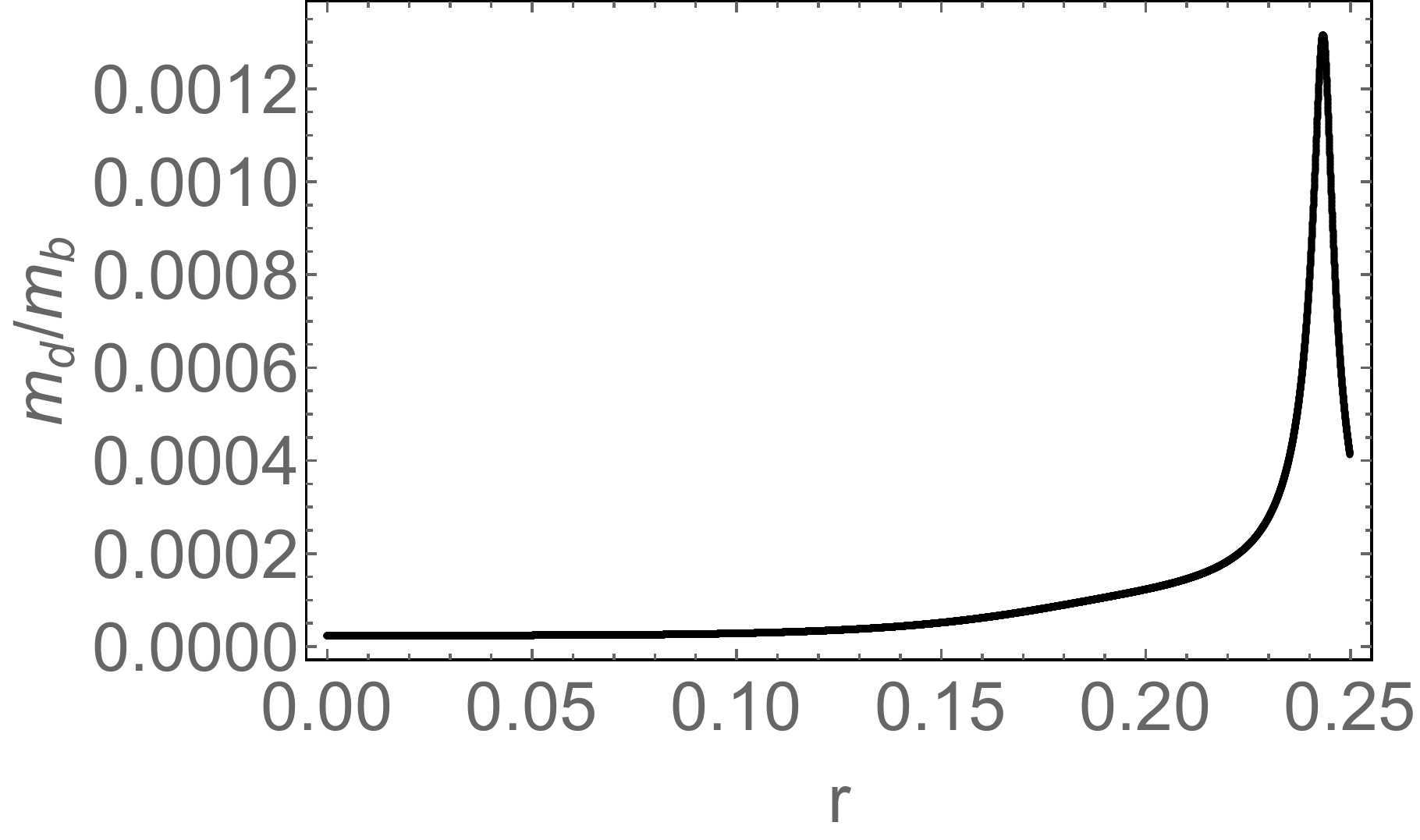}
  \end{minipage}
  \begin{minipage}{0.4\columnwidth}
  \centering
     \includegraphics[scale=0.35]{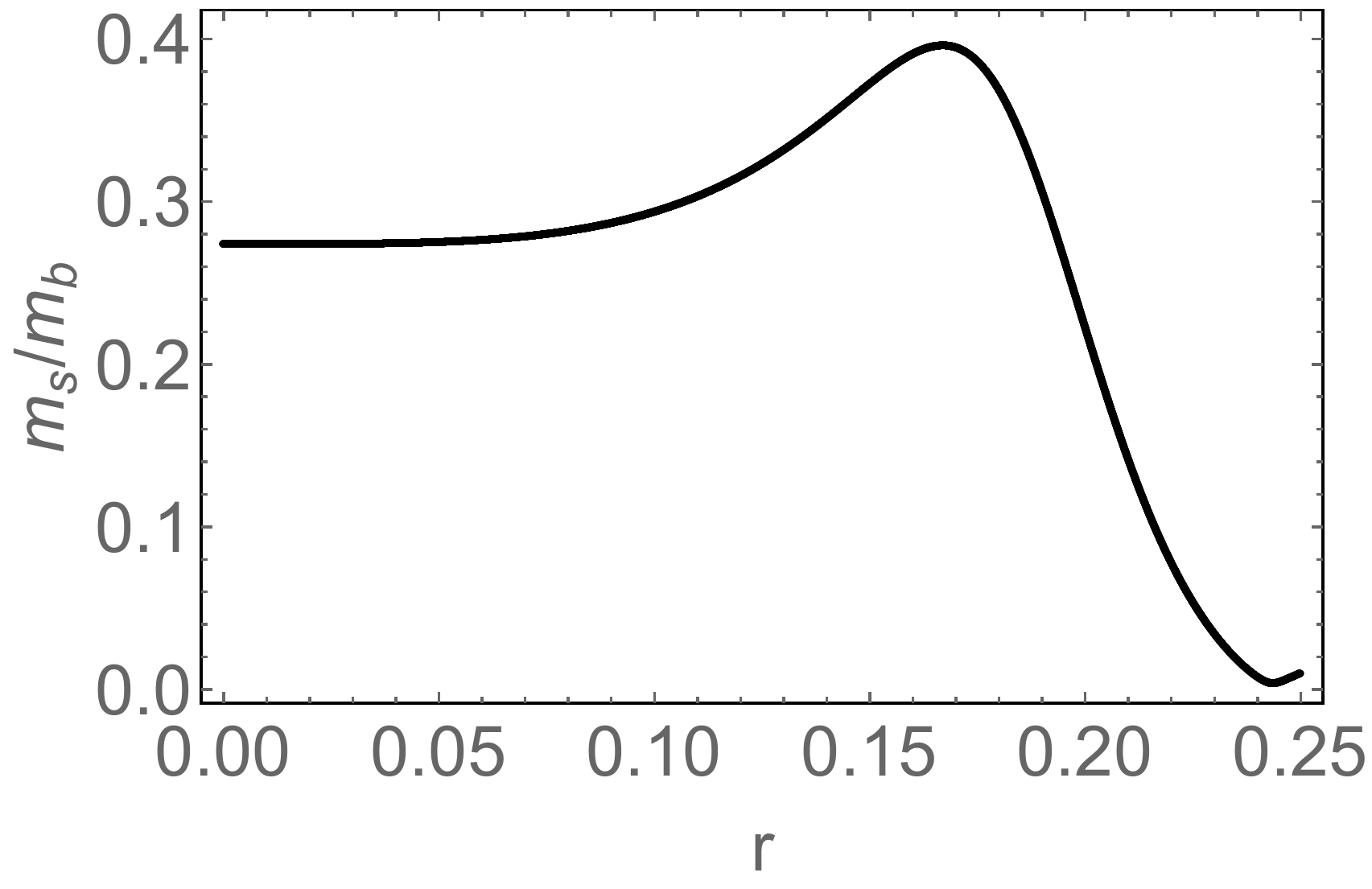}
  \end{minipage}
\caption{The blow-up radius ($r$)-dependence of the mass ratios of the up quark to the top quark $m_u/m_t$, the charm quark to the top quark $m_c/m_t$, the down quark to the bottom quark $m_d/m_b$, and the strange quark to the bottom quark $m_s/m_b$.}
    \label{fig:mass}
\end{figure}

\begin{figure}[H]
\centering
  \begin{minipage}{0.4\columnwidth}
  \centering
     \includegraphics[scale=0.37]{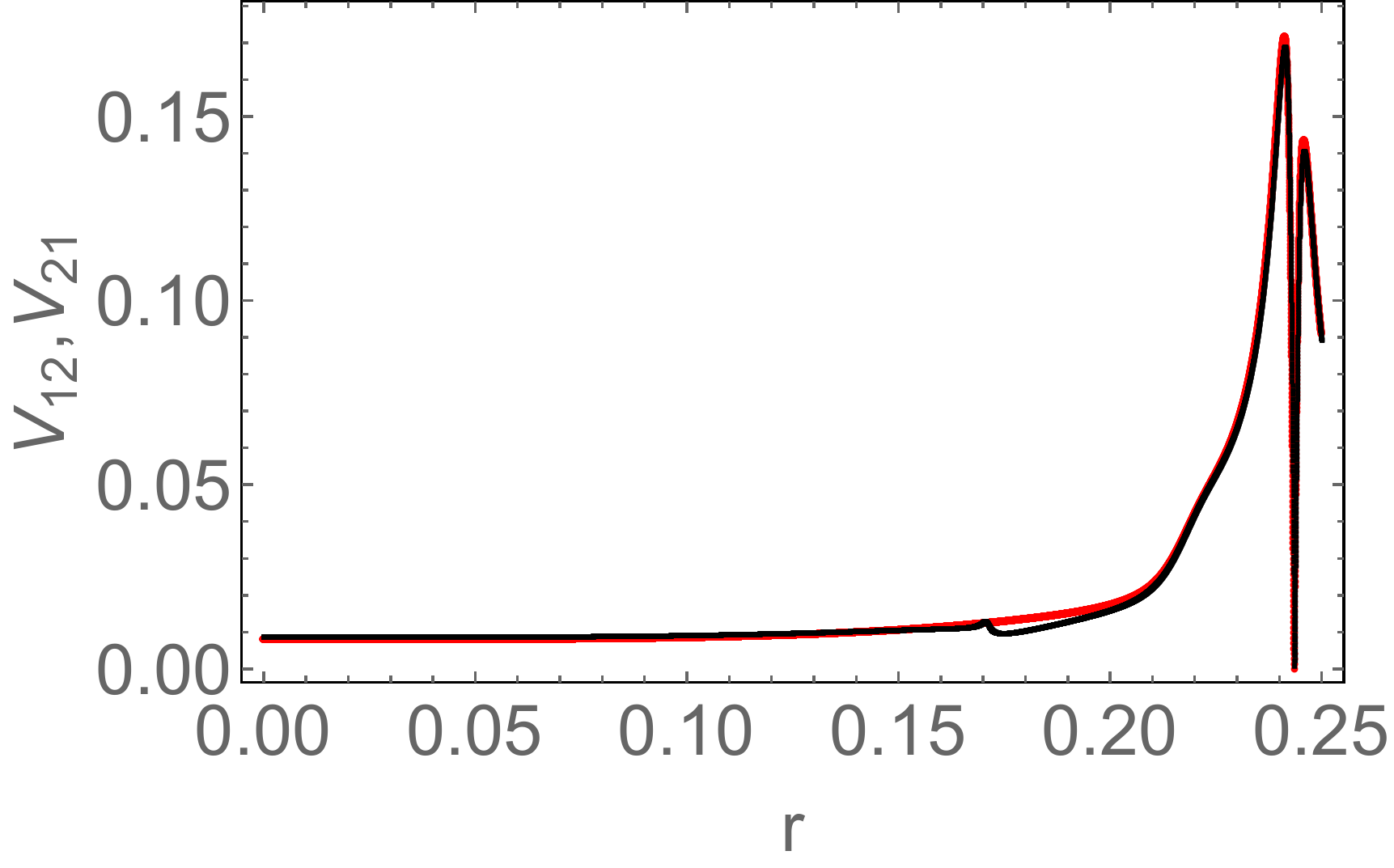}
  \end{minipage}
  \begin{minipage}{0.4\columnwidth}
  \centering
     \includegraphics[scale=0.37]{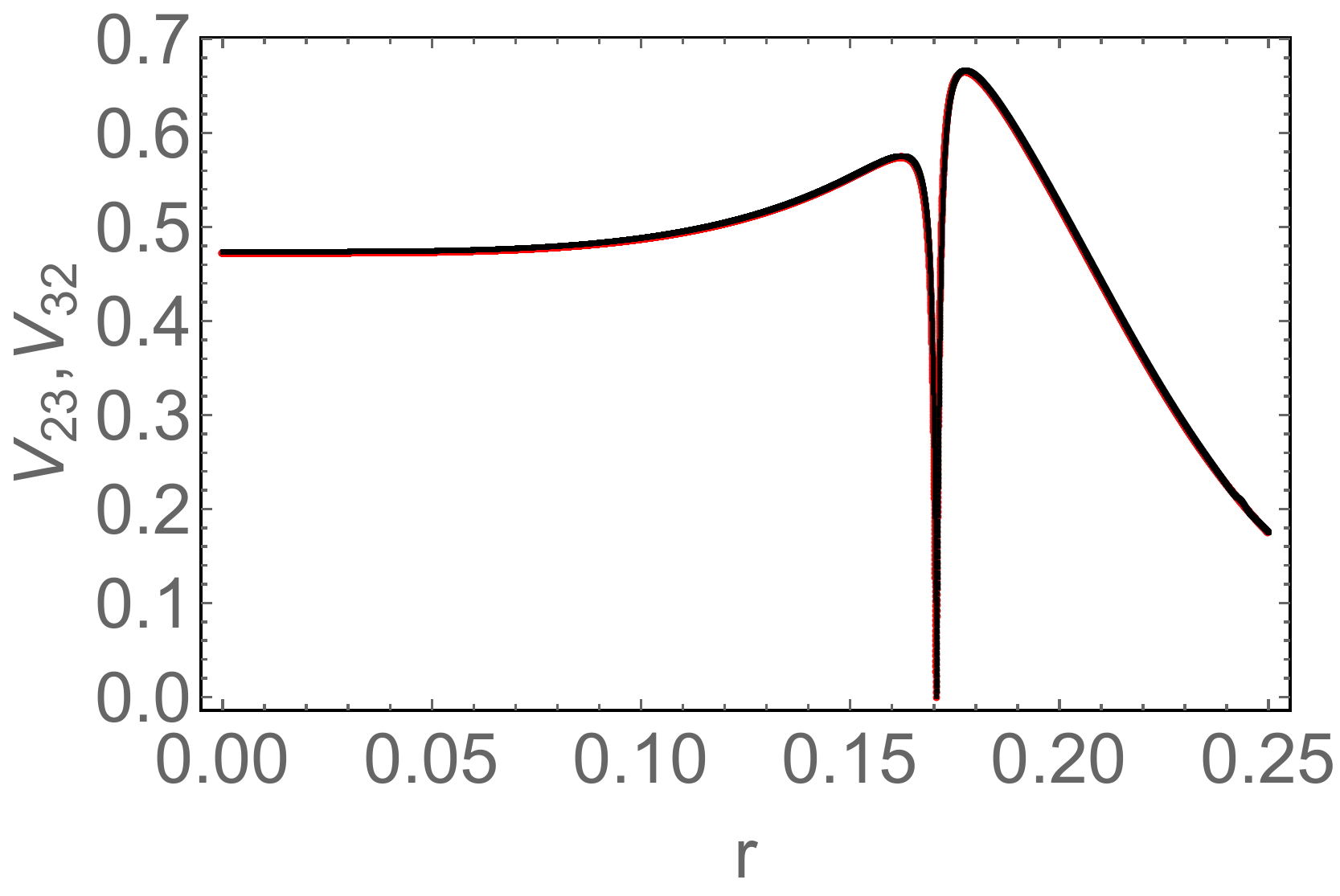}
  \end{minipage} \\
  \begin{minipage}{0.4\columnwidth}
  \centering
     \includegraphics[scale=0.37]{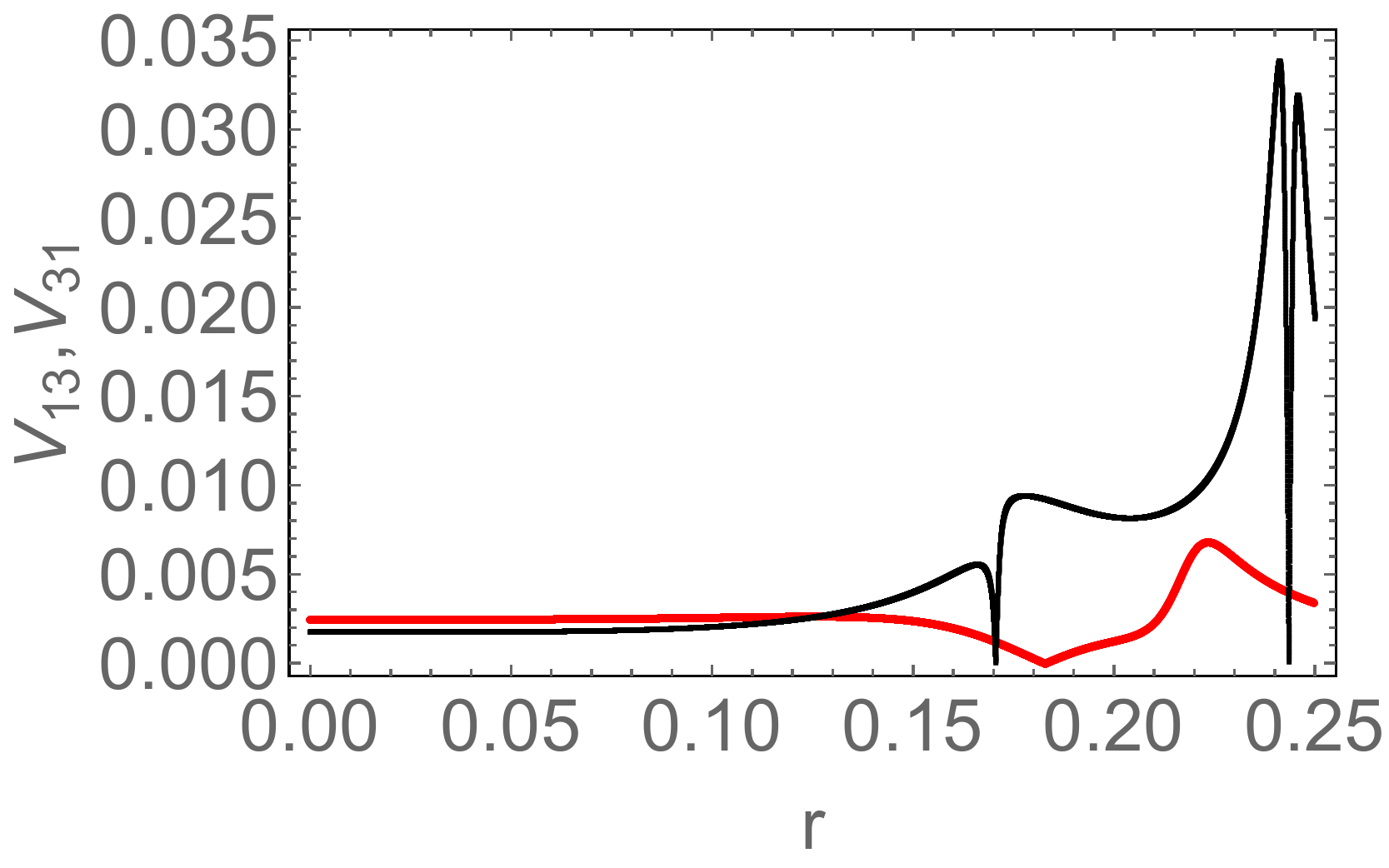}
  \end{minipage}
\caption{The blow-up radius ($r$)-dependence of the CKM mixing angles $V_{12}(V_{21})$, $V_{23}(V_{32})$, and $V_{13}(V_{31})$, as shown in the red (black) lines.}
    \label{fig:CKM}
\end{figure}

Here, we study the model, where the blow-up radii of the four 
fixed points are the same.
We chose $\cos(\pi/6)H_4-\sin(\pi/6)H_3$ and $\sin(\pi/6)H_4+\cos(\pi/6)H_3$ as the directions of the vacuum expectation values (VEVs) for the up- and down-type Higgs fields, respectively. 
The reason to consider two pairs of Higgs doublets is that one can not realize the realistic masses and mixing angles of the quark sector with one pair of Higgs doublets. 
The Yukawa matrices $Y_{K=3}$ and $Y_{K=4}$ are approximated as
\begin{align}
\begin{split}
    Y_{K=3} &= \frac{a}{\sqrt{2}}
    \begin{pmatrix}
    \sqrt{2} \eta_{10000} & \sqrt{2} \eta_{6400} & - \sqrt{2} \eta_{400} \\
    \eta_{4624} & - \eta_{1024} & \eta_{64} \\
    - \eta_{1936} & \eta_{16} & - \eta_{4096}
    \end{pmatrix}
    - \pi^3 r^4 \frac{a}{\sqrt{2}}
    \begin{pmatrix}
    8 \sqrt{2} \eta_{1040} & 96 \sqrt{2} \eta_{800} & - 8 \sqrt{2} \eta_{680} \\
    176 \eta_{704} & 16 \eta_{464} & 128 \eta_{344} \\
    48 \eta_{536} & 128 \eta_{296} & 16 \eta_{176}
    \end{pmatrix}
    ,\\
Y_{K=4} &= \frac{a}{\sqrt{2}}
    \begin{pmatrix}
    \sqrt{2} \eta_{21025} & - \sqrt{2} \eta_{625} & \sqrt{2} \eta_{7225} \\
    \eta_{529} & \eta_{5329} & - \eta_{169} \\
    \eta_{121} & \eta_{121} & \eta_{1}
    \end{pmatrix}
    - \pi^3 r^4 \frac{a}{\sqrt{2}}
    \begin{pmatrix}
    87 \sqrt{2} \eta_{935} & 0 & 63 \sqrt{2} \eta_{575} \\
    28 \eta_{594} & 126 \eta_{359} & 0 \\
    126 \eta_{431} & 28 \eta_{191} & 98 \eta_{71}
    \end{pmatrix},
    \end{split}
    \label{eq:Y4}    
\end{align}
where the overall factor $a=\frac{\left|{\cal N}^{I,5}_1{\cal N}^{J+1,7}_1{\cal N}^{K+1,12}_1\right|}{\left|{\cal N}^{K+1,12}\right|^2}$ and we assume $\pi\tau=5.7i$ and $\eta_n \equiv e^{-5.7n/420}$, originating from 
the approximation of the Jacobi theta function. 
(For more details on the calculation, see, Appendix~\ref{app}.)
By employing the above Yukawa couplings, we calculate the mass ratios of the up/charm quark to the top quark and the down/strange quark to the bottom quark, and Cabibbo-Kobayashi-Maskawa (CKM) matrix.
We show the $r$-dependence of them in FIGs.~\ref{fig:mass} and \ref{fig:CKM}, where the maximum of $r$ is set to be $0.25$ since we use the same $r$ for all the fixed points and the distance to the nearest fixed points is $0.5$.

From FIGs.~\ref{fig:mass} and \ref{fig:CKM}, it is found that the blow-up of the orbifold singularities significantly affects the quark masses and the CKM matrices.
Table~\ref{tb:model1} shows the mass ratios and the CKM matrix 
at $r \approx 0.24\ (\pi^3r^4=0.105)$, and the table also shows the results in the case of $r=0$, that is the  $T^2/\mathbb{Z}_2$ orbifold limit, and the experimental data for magnitudes of CKM elements~\cite{Tanabashi:2018oca} and quark masses evaluated at the so-called GUT scale $2.0\times 10^{16}\,{\rm GeV}$~\cite{Xing:2008}. 
Thus, we find that the blow-up of the orbifold singularities makes the values approach the observed values up to $O(1)$.


\begin{table}[H]
\centering
\scalebox{0.9}{
    \begin{tabular}{|c|c|c|c|} \hline 
    \par & Pure $T^2/\mathbb{Z}_2$ ($r=0$) & Blow-ups of $T^2/\mathbb{Z}_2$ ($r \approx 0.24$) & Observed values \\ \hline
	$(m_u, m_c, m_t)/m_t$ & $(1.5 \times 10^{-4}, 9.6 \times 10^{-2}, 1)$ & $(1.2 \times 10^{-5}, 2.3 \times 10^{-2}, 1)$ & $(6.5 \times 10^{-6}, 3.2 \times 10^{-3}, 1)$ \\ \hline
	$(m_d, m_s, m_b)/m_b$ & $(2.5 \times 10^{-5}, 2.7 \times 10^{-1}, 1)$ & $(9.7 \times 10^{-4}, 6.3 \times 10^{-3}, 1)$ & $(1.1 \times 10^{-3}, 2.2 \times 10^{-2}, 1)$ \\ \hline
	$|V_{CKM}|$
	&$\begin{pmatrix}
	1.0 & 0.0082 & 0.0022 \\
	0.0084 & 0.88 & 0.47 \\
	0.0017 & 0.47 & 0.88
	\end{pmatrix}$
	&$\begin{pmatrix}
	0.99 & 0.17 & 0.0043 \\
	0.17 & 0.96 & 0.22 \\
	0.034 & 0.22 & 0.98
	\end{pmatrix}$
	&$\begin{pmatrix}
	0.97 & 0.22 & 0.0037 \\
	0.22 & 0.97 & 0.042 \\
	0.0090 & 0.041 & 1.0
	\end{pmatrix}$
	\\ \hline
	\end{tabular}}
	\caption{
		The mass ratios and CKM matrices for the blow-up radius $r=0$ (the pure $T^2/\mathbb{Z}_2$ orbifold), $r \approx 0.24$, and the observed values, where we use the GUT scale running masses~\cite{Xing:2008}.}
	\label{tb:model1}
\end{table}

%%%%%%%%%%%%%%%%%%%%%%%%%%%%%%%%%%%%%%%%%%%%%%%%%%%%%%%%%%%%%%%%%%%%%%%%%%%%%%%%%%%%%%%%%%%%%%%%%%%%%%%%%%%%%%%%%%%%%%%%%%%%%%%%%%%%
\subsection{Model 2}
\label{sec:3_2}

So far, we have taken into account the blow-up of all the fixed points, each with the same 
blow-up radius, but there is no reason to take the same values of blow-up radii 
which can be determined by the dynamics of blow-up modes, i.e. the blow-up moduli stabilization. 

\begin{figure}[H]
\centering
  \begin{minipage}{0.4\columnwidth}
  \centering
     \includegraphics[scale=0.32]{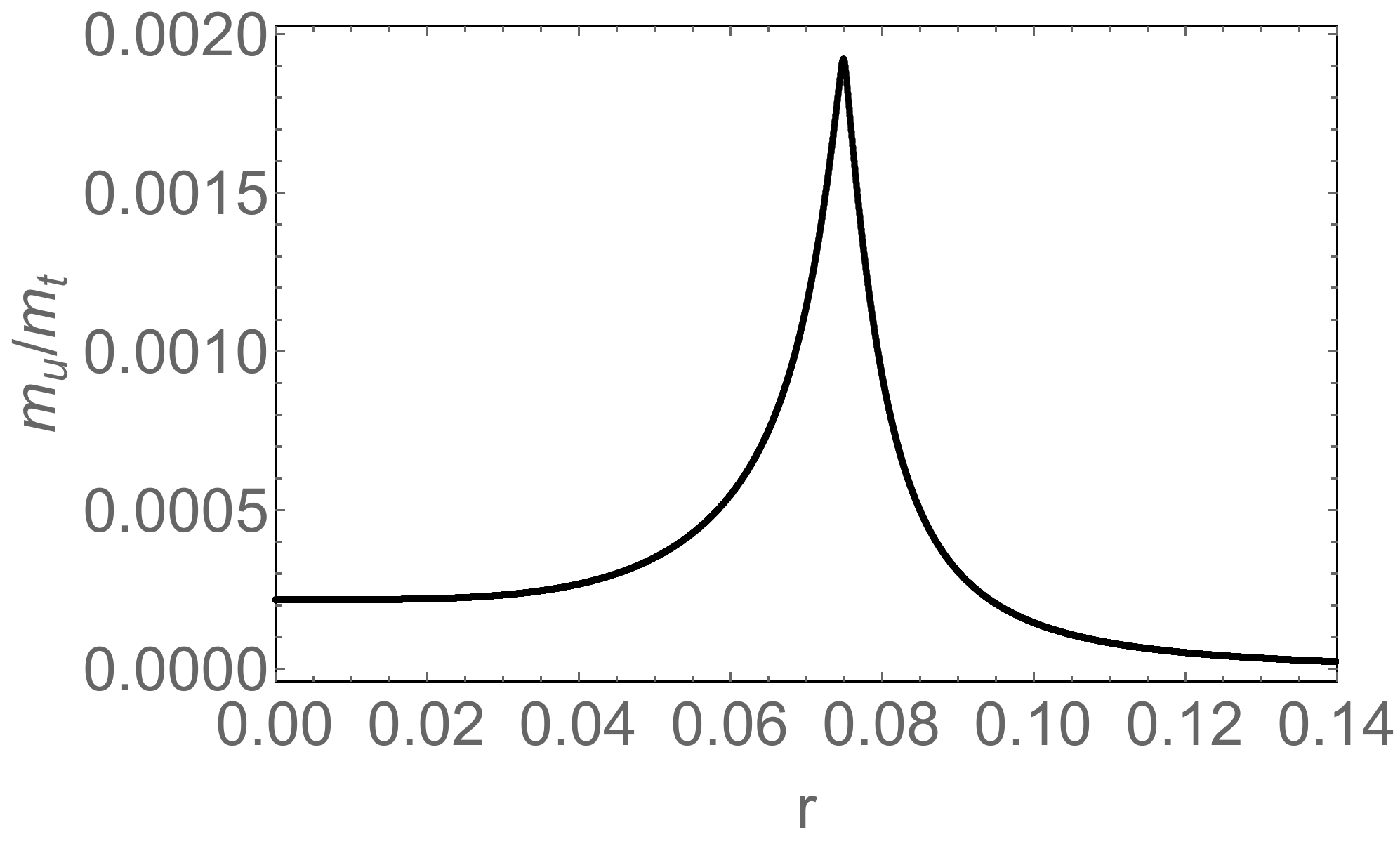}
  \end{minipage}
  \begin{minipage}{0.4\columnwidth}
  \centering
     \includegraphics[scale=0.3]{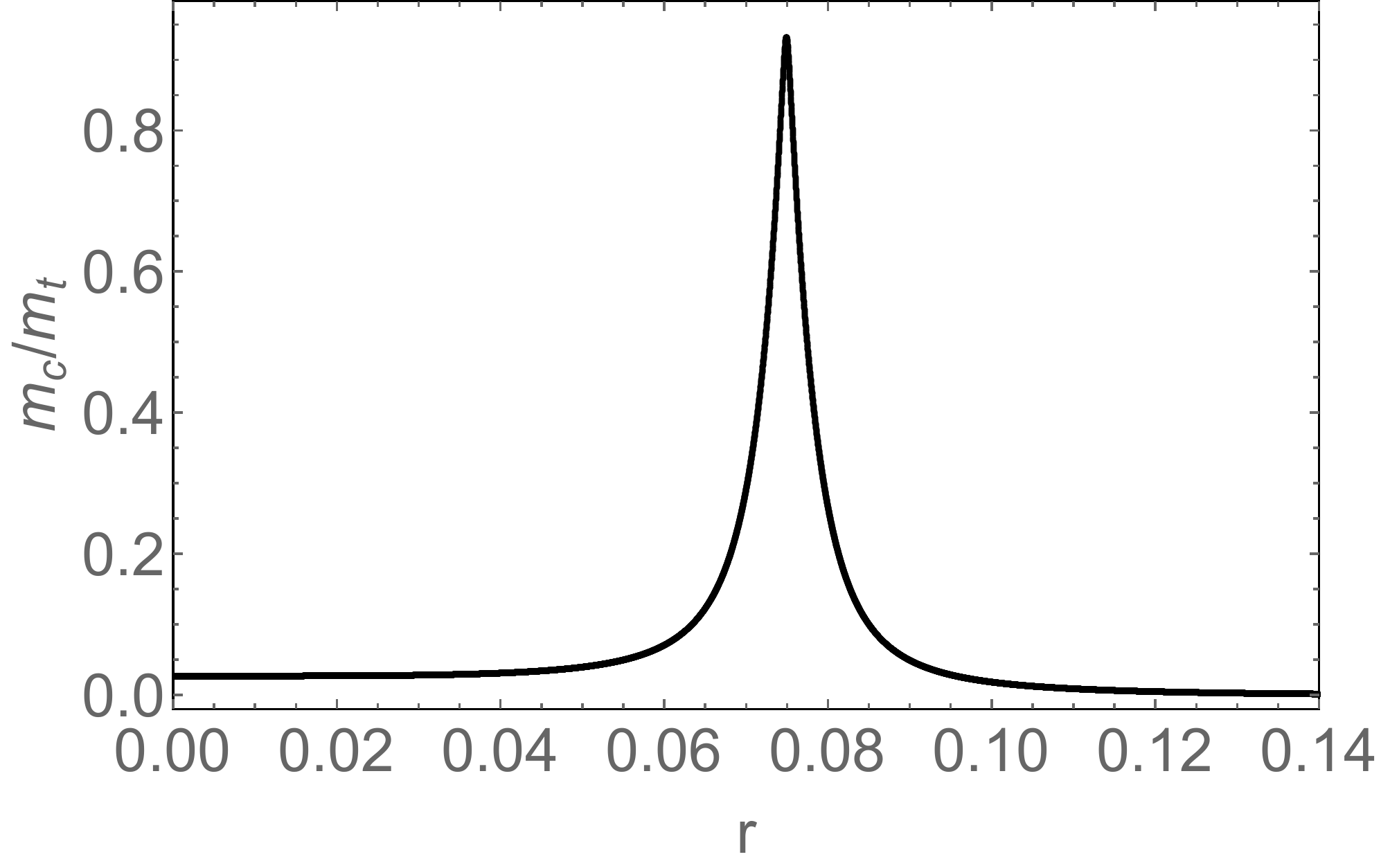}
  \end{minipage} \\
  \begin{minipage}{0.4\columnwidth}
  \centering
     \includegraphics[scale=0.32]{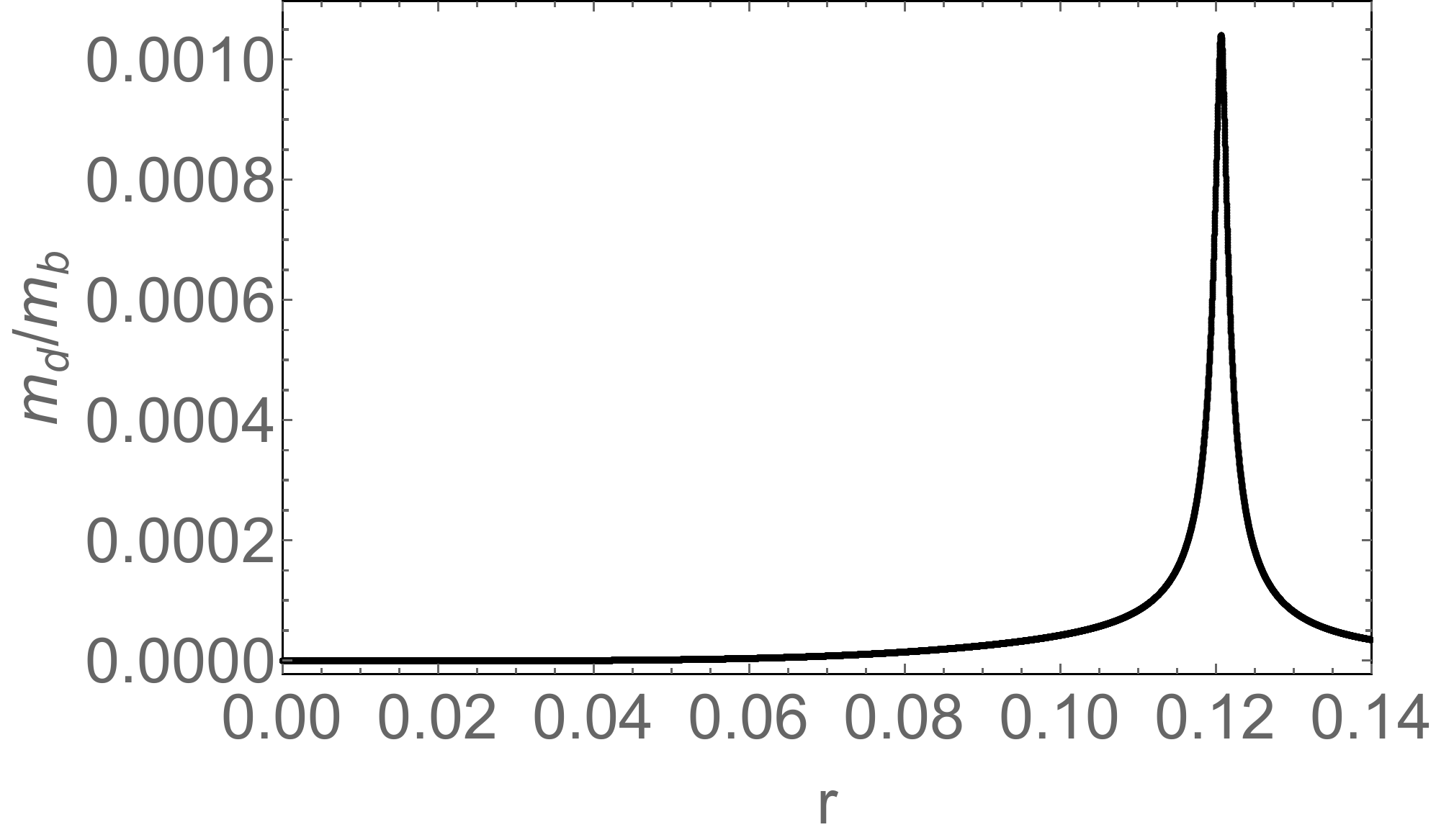}
  \end{minipage}
  \begin{minipage}{0.4\columnwidth}
  \centering
     \includegraphics[scale=0.3]{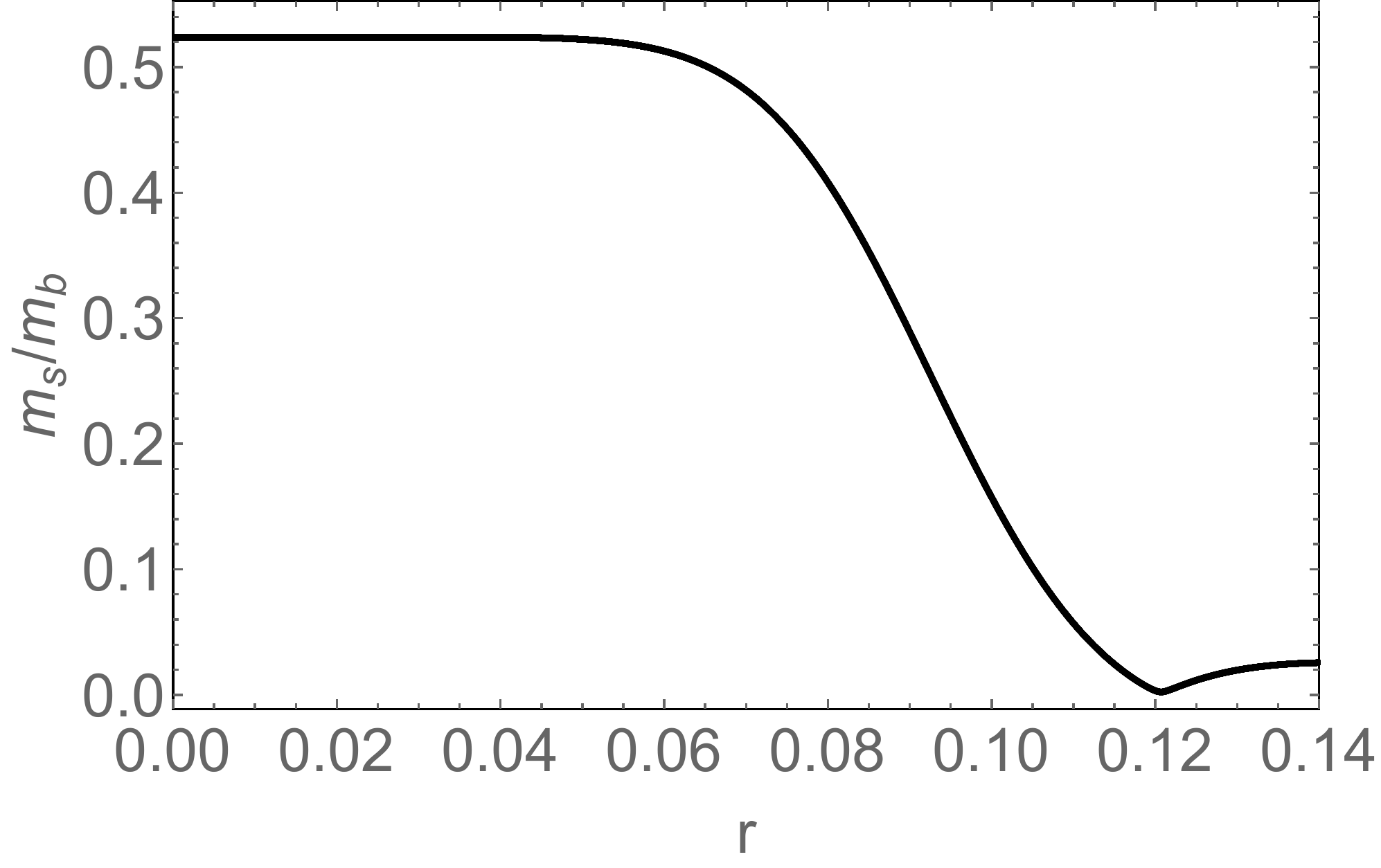}
  \end{minipage}
\caption{The blow-up radius ($r$)-dependence of the mass ratios of the up quark to the top quark $m_u/m_t$, the charm quark to the top quark $m_c/m_t$, the down quark to the bottom quark $m_d/m_b$, and the strange quark to the bottom quark $m_s/m_b$.}
    \label{fig:mass2}
\end{figure}

\begin{figure}[H]
\centering
  \begin{minipage}{0.4\columnwidth}
  \centering
     \includegraphics[scale=0.35]{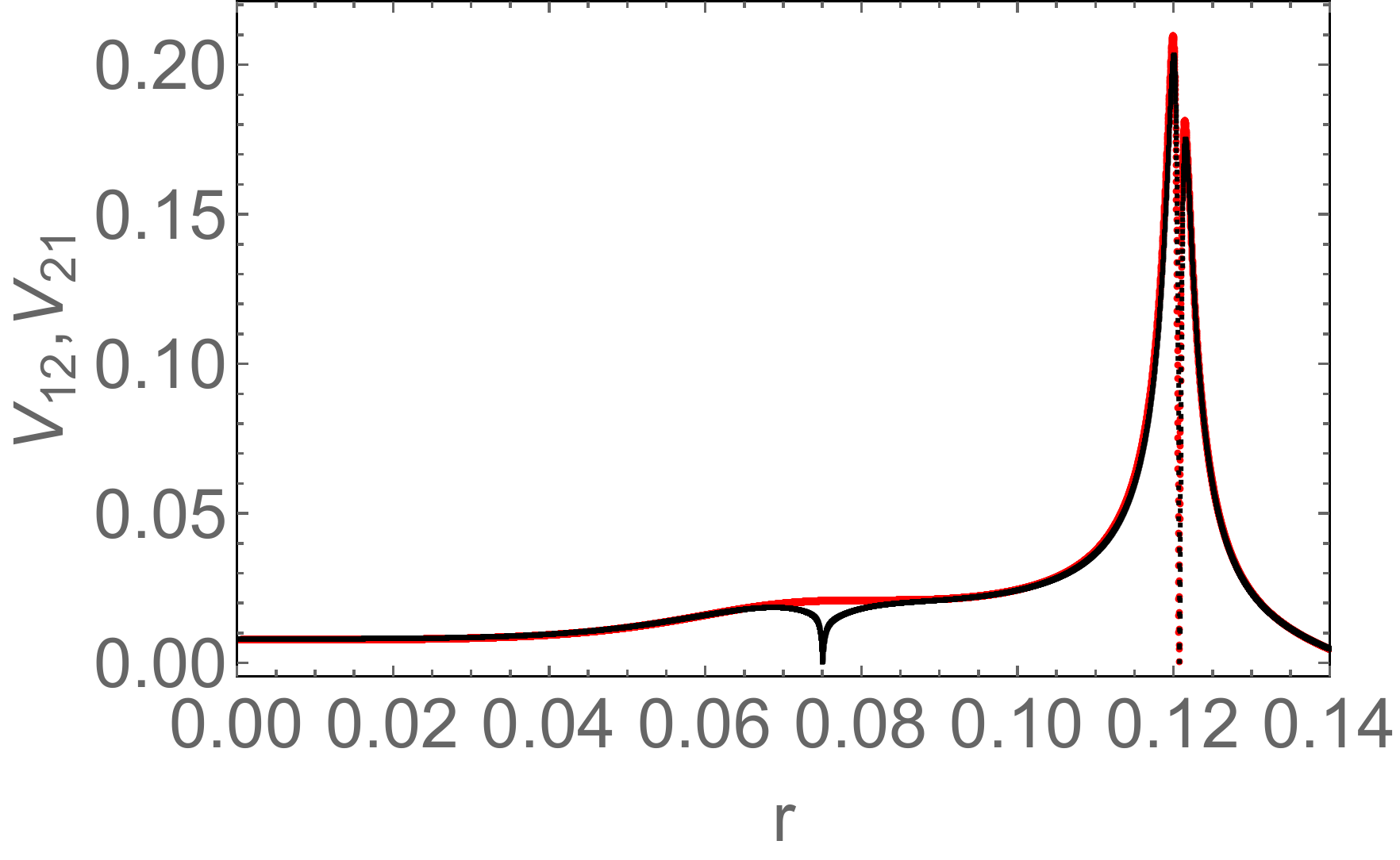}
  \end{minipage}
  \begin{minipage}{0.4\columnwidth}
  \centering
     \includegraphics[scale=0.35]{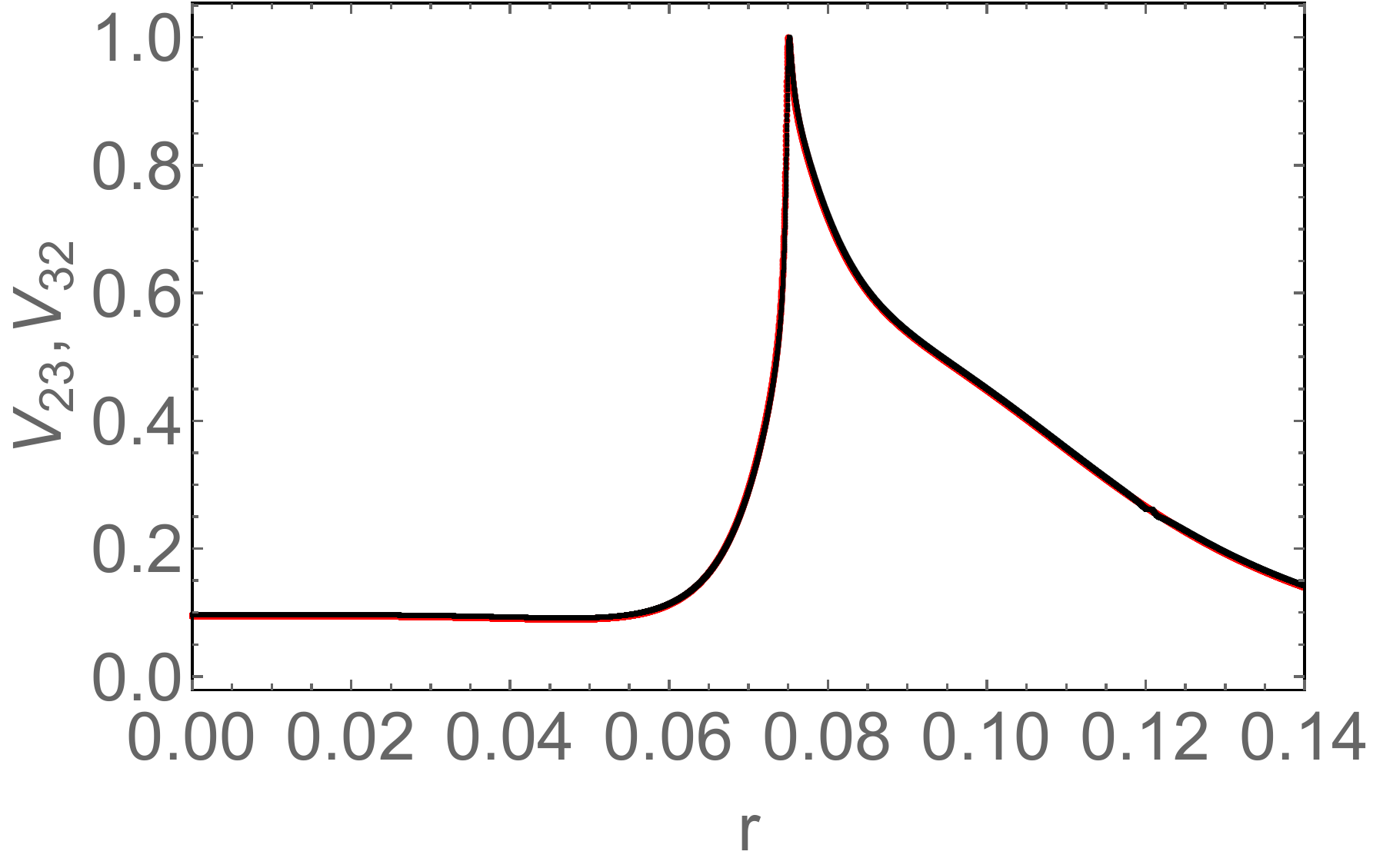}
  \end{minipage} \\
  \begin{minipage}{0.4\columnwidth}
  \centering
     \includegraphics[scale=0.35]{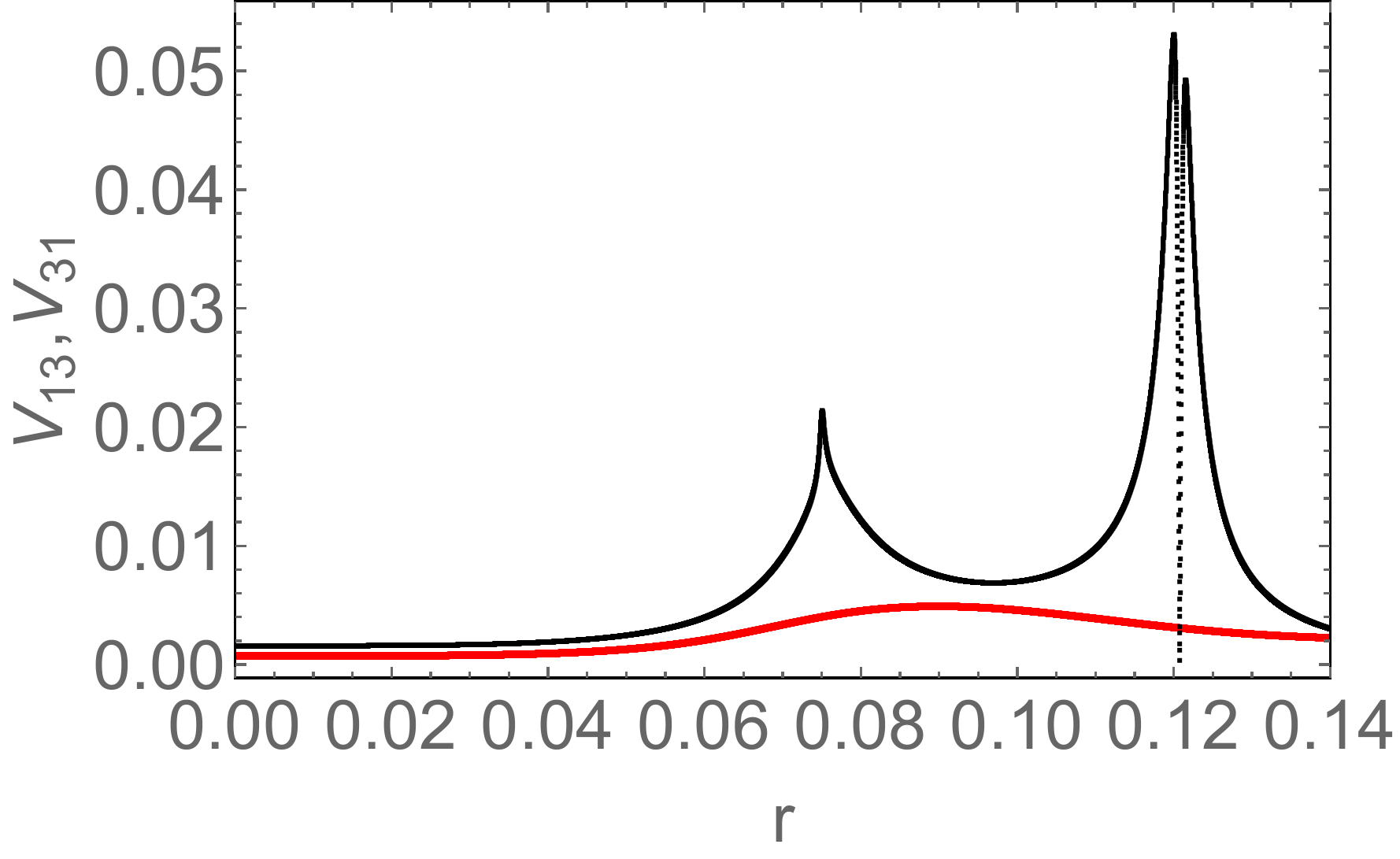}
  \end{minipage}
\caption{The blow-up radius ($r$)-dependence of the CKM mixing angles $V_{12}(V_{21})$, $V_{23}(V_{32})$, and $V_{13}(V_{31})$, as shown in the red (black) lines.}
    \label{fig:CKM2}
\end{figure}

In this section, we examine the case where the blow-up radius at $z_I=\tau/2$ fixed point $r_{\tau/2}$ is different from the others, namely $r=r_0=r_{1/2}=r_{(\tau+1)/2}$, 
where $r_0$, $r_{1/2}$, $r_{\tau/2}$ $r_{(\tau+1)/2}$ denote the blow-up radii at the fixed 
points $z_I=0, 1/2, \tau/2, (\tau+1)/2$, respectively. 
To simplify our analysis, we focus on the model with one pair of Higgs doublets, where 
the up- and down-type Higgs fields are identified with $H_u=H_4$ and $H_d=H_3$. 
The Yukawa coupling $Y_{IJK}$ in Eq.~(\ref{eq:Yukawa}) is calculated by using 
the formulae in Appendix~\ref{app}. 
In a way similar to the previous section, we draw the mass ratios and magnitudes of CKM elements 
as a function of blow-up radius $r$ in FIGs.~\ref{fig:mass2} and~\ref{fig:CKM2}, 
in which we assume 
\begin{align}
\tau=1.8i, \qquad
r_{\tau/2} = 2.6 \times r.
\end{align}
It indicates that the maximum value of $r$ is $0.139$, otherwise 
two blow-up regions associated with the fixed points $z=\tau/2$ and $z=(\tau +1)/2$ 
overlap with each other.

Table~\ref{tb:model2} shows 
the predictions of the mass ratios and the CKM mixing angles for both the magnetized orbifold 
model ($r=0$) and the magnetized blow-up model ($r =0.12$). 
It turns out that the rank of down-type quark mass matrix is 2 in the toroidal orbifold model, 
but blowing up the fixed points leads to the rank 3 mass matrix and more realistic 
flavor structure.

\begin{table}[H]
\centering
\scalebox{0.9}{
    \begin{tabular}{|c|c|c|c|} \hline 
    \par & Pure $T^2/\mathbb{Z}_2$ ($r=0$) & Blow-ups of $T^2/\mathbb{Z}_2$ ($r =0.12$) & Observed values \\ \hline
	$(m_u, m_c, m_t)/m_t$ & $(2.2 \times 10^{-4}, 2.7 \times 10^{-2}, 1)$ & $(5.3 \times 10^{-5}, 5.2 \times 10^{-3}, 1)$ & $(6.5 \times 10^{-6}, 3.2 \times 10^{-3}, 1)$ \\ \hline
	$(m_d, m_s, m_b)/m_b$ & $(0, 5.2 \times 10^{-1}, 1)$ & $(7.6 \times 10^{-4}, 3.9 \times 10^{-3}, 1)$ & $(1.1 \times 10^{-3}, 2.2 \times 10^{-2}, 1)$ \\ \hline
	$|V_{CKM}|$
	&$\begin{pmatrix}
	1.0 & 0.0080 & 0.00077 \\
	0.0084 & 1.0 & 0.096 \\
	0.0015 & 0.0096 & 1.0
	\end{pmatrix}$
	&$\begin{pmatrix}
	0.98 & 0.21 & 0.0032 \\
	0.20 & 0.94 & 0.27 \\
	0.053 & 0.26 & 0.96
	\end{pmatrix}$
	&$\begin{pmatrix}
	0.97 & 0.22 & 0.0037 \\
	0.22 & 0.97 & 0.042 \\
	0.0090 & 0.041 & 1.0
	\end{pmatrix}$
	\\ \hline
	\end{tabular}}
	\caption{
		The mass ratios and CKM matrices for the blow-up radius $r=0$ (the pure $T^2/\mathbb{Z}_2$ orbifold), $r =0.12$, and the observed values, where we use the GUT scale running masses~\cite{Xing:2008}. }
	\label{tb:model2}
\end{table}

%%%%%%%%%%%%%%%%%%%%%%%%%%%%%%%%%%%%%%%%%%%%%%%%%%%%%%%%%%%%%%%%%%%%%%%%%%%%%%%%%%%%%%%%%%%%%%%%%%%%%%%%%%%%%%%%%%%%%%%%%%%%%%%%%%%%

\section{Conclusions}
\label{sec:con}

We have examined the flavor structure of magnetized $T^2/\mathbb{Z}_2$ 
blow-up models, where the orbifold fixed points are replaced by a part of sphere 
proposed by the method in Ref.~\cite{Kobayashi:2019}. 
The flavor structure as well as the selection rules of Yukawa couplings among chiral zero-modes 
are different from the toroidal orbifold results, due to the blowing up the orbifold 
singularities as demonstrated in a concrete model in Section~\ref{sec:3}. 
Interestingly, the elements of the CKM matrix and mass hierarchy of the quark sector are well 
consistent with the observed data at certain values of the blow-up radii, rather than the toroidal orbifold result. 
Furthermore, such a realistic setup can be realized by a few number of parameters, although in toroidal 
orbifold case, several pairs of Higgs doublets are required to explain the observed flavor structure and mass hierarchy of the quark sector.

In this paper, we have assumed that the blow-up modes determining the blow-up radii are 
considered just the parameter, but it should be taken as a dynamical degree of freedom. 
It is interesting to discuss the dynamics of  the blow-up modes such that their vacuum 
expectation values lead to the observed data. 
We have focused on the simplest $T^2/\mathbb{Z}_2$ case, but our analysis can be 
generalized to resolutions of $T^2/\mathbb{Z}_N$ orbifolds \footnote{See the models at the orbifold limit,  Refs.~\cite{Abe:2013bca,Abe:2014noa}.}. 
Another possible generalization of our work is to apply our method to the resolutions of higher dimensional toroidal orbifolds \footnote{For example, it is interesting to discuss the wavefunctions on blow-ups of the $T^4/\mathbb{Z}_2$ orbifold, using the numerical metric of K3 surface~\cite{Headrick:2005ch} and the selection rules of matter wavefunctions along the line of Refs.\cite{Burgess:2008ri,Maharana:2011wx}.}.
We hope to report on these subjects in the future.

%%%%%%%%%%%%%%%%%%%%%%%%%%%%%%%%%%%%%%%%%%%%%%%%%%%%%%%%%%%%%%%%%%%%%%%%%%%%%%%%%%%%%%%%%%%%%%%%%%%%%%%%%%%%%%%%%%%%%%%%%%%%%%%%%%%%

\section*{Acknowledgments}
T. K. was supported in part by MEXT KAKENHI Grant Number JP19H04605. 
H. O. was supported by Grant-in-Aid for JSPS Research Fellows No. 19J00664.

%%%%%%%%%%%%%%%%%%%%%%%%%%%%%%%%%%%%%%%%%%%%%%%%%%%%%%%%%%%%%%%%%%%%%%%%%%%%%%%%%%%%%%%%%%%%%%%%%%%%%%%%%%%%%%%%%%%%%%%%%%%%%%%%%%%%

\appendix

\section{Correction terms of Yukawa coupling}
\label{app}
Here, we show the explicit calculation of Eq. (\ref{eq:second}) at each singularity as follows:
\begin{align}
&\int_{\left|z\right| \leq r} dzd\bar{z}\ \phi^{I,5}_{T^2/\mathbb{Z}_2^+}(z) \phi^{J+1,7}_{T^2/\mathbb{Z}_2^-}(z) \left(\phi^{K+1,12}_{T^2/\mathbb{Z}_2^-}(z)\right)^{\ast} \notag \\
&\simeq \frac{1}{2\pi} \left(\frac{\pi r^2}{2}\right)^2 \phi^{I,5}_{T^2/\mathbb{Z}_2^+}(0) \phi'^{J+1,7}_{T^2/\mathbb{Z}_2^-} (0) \left(\phi'^{k,M}_{T^2/\mathbb{Z}_2^-}(0)\right)^{\ast} \notag \\
&= \sqrt{2} \pi^3 r^4 a \sum_{l,m,n} \left(J+1+7m\right)\left(K+1+12n\right) e^{-\pi {\rm Im}\tau \left[5\left(\frac{I}{5}+l\right)^2+7\left(\frac{J+1}{7}+m\right)^2+12\left(\frac{K+1}{12}+n\right)^2\right]} \notag \\
%&\simeq \sqrt{2} \pi^3 r^4 a \left(J+1\right)\left(K+1\right) e^{i\pi \tau \left[\frac{I^2}{5}+\frac{\left(J+1\right)^2}{7}+\frac{\left(K+1\right)^2}{12}\right]} \notag \\
&\simeq \sqrt{2} \pi^3 r^4 a \left(J+1\right)\left(K+1\right) e^{-\pi {\rm Im}\tau \frac{84I^2+60\left(J+1\right)^2+35\left(K+1\right)^2}{420}} \notag \\
&= \sqrt{2} \pi^3 r^4 a \left(J+1\right)\left(K+1\right) \eta_{84I^2+60\left(J+1\right)^2+35\left(K+1\right)^2} 
,
\label{0}
\end{align}
\begin{align}
&\int_{\left|z-\frac{1}{2}\right| \leq r} dzd\bar{z}\ \phi^{I,5}_{T^2/\mathbb{Z}_2^+}(z) \phi^{J+1,7}_{T^2/\mathbb{Z}_2^-}(z) \left(\phi^{K+1,12}_{T^2/\mathbb{Z}_2^-}(z)\right)^{\ast} \notag \\
&\simeq \frac{1}{2\pi} \left(\frac{\pi r^2}{2}\right)^2 \phi^{I,5}_{T^2/\mathbb{Z}_2^+}(\frac{1}{2}) \phi'^{J+1,7}_{T^2/\mathbb{Z}_2^-} (\frac{1}{2}) \left(\phi'^{k,M}_{T^2/\mathbb{Z}_2^-}(\frac{1}{2})\right)^{\ast} \notag \\
&= \sqrt{2} \pi^3 r^4 a \sum_{l,m,n} \left(J+1+7m\right)\left(K+1+12n\right) \notag \\
&\times e^{-\pi {\rm Im}\tau \left[5\left(\frac{I}{5}+l\right)^2+7\left(\frac{J+1}{7}+m\right)^2+12\left(\frac{K+1}{12}+n\right)^2\right]} e^{\pi i\left[\left(I+5l\right)+\left(J+1+7m\right)-\left(K+1+12n\right)\right]} \notag \\
%&\simeq \sqrt{2} \pi^3 r^4 a \left(J+1\right)\left(K+1\right) e^{i\pi \tau \left[\frac{I^2}{5}+\frac{\left(J+1\right)^2}{7}+\frac{\left(K+1\right)^2}{12}\right]} e^{\pi i\left(I+J-K\right)} \notag \\
&\simeq \sqrt{2} \pi^3 r^4 a \left(J+1\right)\left(K+1\right) e^{-\pi {\rm Im}\tau \frac{84I^2+60\left(J+1\right)^2+35\left(K+1\right)^2}{420}} e^{\pi i\left(I+J-K\right)} \notag \\
&= \sqrt{2} \pi^3 r^4 a \left(J+1\right)\left(K+1\right) \eta_{84I^2+60\left(J+1\right)^2+35\left(K+1\right)^2} e^{\pi i\left(I+J-K\right)} 
,
\label{1/2} 
\end{align}
\begin{align}
&\int_{\left|z-\frac{\tau}{2}\right| \leq r} dzd\bar{z}\ \phi^{I,5}_{T^2/\mathbb{Z}_2^+}(z) \phi^{J+1,7}_{T^2/\mathbb{Z}_2^-}(z) \left(\phi^{K+1,12}_{T^2/\mathbb{Z}_2^-}(z)\right)^{\ast} \notag \\
&\simeq \frac{1}{2\pi} \left(\frac{\pi r^2}{2}\right)^2 \phi^{I,5}_{T^2/\mathbb{Z}_2^+}(\frac{\tau}{2}) \phi'^{J+1,7}_{T^2/\mathbb{Z}_2^-} (\frac{\tau}{2}) \left(\phi'^{k,M}_{T^2/\mathbb{Z}_2^-}(\frac{\tau}{2})\right)^{\ast} \notag \\
&= \sqrt{2} \pi^3 r^4 a \sum_{l,m,n} \left(J+1+7m\right)\left(K+1+12n\right) e^{-\pi {\rm Im}\tau \left[5\left(\frac{I}{5}+l+\frac{1}{2}\right)^2+7\left(\frac{J+1}{7}+m+\frac{1}{2}\right)^2+12\left(\frac{K+1}{12}+n+\frac{1}{2}\right)^2\right]} \notag \\
%&\simeq \sqrt{2} \pi^3 r^4 a \left(7-\left(J+1\right)\right)\left(12-\left(K+1\right)\right) e^{- 5.7 \left[6-\frac{I\left(5-I\right)}{5}-\frac{\left(J+1\right)\left(7-\left(J+1\right)\right)}{7}-\frac{\left(K+1\right)\left(12-\left(K+1\right)\right)}{12}\right]} \notag \\
&\simeq \sqrt{2} \pi^3 r_3^4 a \left(7-\left(J+1\right)\right)\left(12-\left(K+1\right)\right) e^{-\pi {\rm Im}\tau \frac{2520-84I\left(5-I\right)-60\left(J+1\right)\left(7-\left(J+1\right)\right)-35\left(K+1\right)\left(12-\left(K+1\right)\right)}{420}} \notag \\
&= \sqrt{2} \pi^3 r^4 a \left(7-\left(J+1\right)\right)\left(12-\left(K+1\right)\right) \eta_{2520-84I\left(5-I\right)-60\left(J+1\right)\left(7-\left(J+1\right)\right)-35\left(K+1\right)\left(12-\left(K+1\right)\right)}
,
\label{tau} 
\end{align}
\begin{align}
&\int_{\left|z-\frac{\tau+1}{2}\right| \leq r} dzd\bar{z}\ \phi^{I,5}_{T^2/\mathbb{Z}_2^+}(z) \phi^{J+1,7}_{T^2/\mathbb{Z}_2^-}(z) \left(\phi^{K+1,12}_{T^2/\mathbb{Z}_2^-}(z)\right)^{\ast} \notag \\
&\simeq \frac{1}{2\pi} \left(\frac{\pi r^2}{2}\right)^2 \phi'^{I,5}_{T^2/\mathbb{Z}_2^+}(\frac{\tau+1}{2}) \phi^{J+1,7}_{T^2/\mathbb{Z}_2^-} (\frac{\tau+1}{2}) \left(\phi'^{k,M}_{T^2/\mathbb{Z}_2^-}(\frac{\tau+1}{2})\right)^{\ast} \notag \\
&= \sqrt{2} \pi^3 r^4 a \sum_{l,m,n} \left(I+5l\right)\left(K+1+12n\right) \notag \\
&\times e^{-\pi {\rm Im}\tau \left[5\left(\frac{I}{5}+l+\frac{1}{2}\right)^2+7\left(\frac{J+1}{7}+m+\frac{1}{2}\right)^2+12\left(\frac{K+1}{12}+n+\frac{1}{2}\right)^2\right]} e^{\pi i\left[\left(I+5l\right)+\left(J+1+7m\right)-\left(K+1+12n\right)\right]} \notag \\
%&\simeq \sqrt{2} \pi^3 r^4 a \left(5-I\right)\left(12-\left(K+1\right)\right) e^{- 5.7 \left[6-\frac{I\left(5-I\right)}{5}-\frac{\left(J+1\right)\left(7-\left(J+1\right)\right)}{7}-\frac{\left(K+1\right)\left(12-\left(K+1\right)\right)}{12}\right]} e^{\pi i\left(I+J-K\right)} \notag \\
&\simeq \sqrt{2} \pi^3 r^4 a \left(5-I\right)\left(12-\left(K+1\right)\right) e^{-\pi {\rm Im}\tau \frac{2520-84I\left(5-I\right)-60\left(J+1\right)\left(7-\left(J+1\right)\right)-35\left(K+1\right)\left(12-\left(K+1\right)\right)}{420}} e^{\pi i\left(I+J-K\right)} \notag \\
&= \sqrt{2} \pi^3 r^4 a \left(5-I\right)\left(12-\left(K+1\right)\right) \eta_{2520-84I\left(5-I\right)-60\left(J+1\right)\left(7-\left(J+1\right)\right)-35\left(K+1\right)\left(12-\left(K+1\right)\right)} e^{\pi i\left(I+J-K\right)}, 
\label{tau1}
\end{align}
where the overall factor $a=\frac{\left|{\cal N}^{I,5}_1{\cal N}^{J+1,7}_1{\cal N}^{K+1,12}_1\right|}{\left|{\cal N}^{K+1,12}\right|^2}$ 
%and we assume $\pi\tau=5.7i$ and $\eta_n \equiv e^{-5.7n/420}$.
and we asuume ${\rm Re}\tau=0$ and approximate the Jacobi theta function by $\eta_n \equiv e^{- n\pi {\rm Im}\tau /420}$.
Note that we need to multiply $1/\sqrt{2}$ further if $I=0$.

%%%%%%%%%%%%%%%%%%%%%%%%%%%%%%%%%%%%%%%%%%%%%%%%%%%%%%%%%%%%%%%%%%%%%%%%%%%%%%%%%%%%%%%%%%%%%%%%%%%%%%%%%%%%%%%%%%%%%%%%%%%%%%%%%%%%


\begin{thebibliography}{99}



%\cite{Abe:2008fi}
\bibitem{Abe:2008fi} 
  H.~Abe, T.~Kobayashi and H.~Ohki,
  %``Magnetized orbifold models,''
  JHEP {\bf 0809}, 043 (2008)
% doi:10.1088/1126-6708/2008/09/043
  [arXiv:0806.4748 [hep-th]].
%%CITATION = doi:10.1088/1126-6708/2008/09/043;%%

%\cite{Abe:2009}
\bibitem{Abe:2009}
  H.~Abe, K.~S.~Choi, T.~Kobayashi and H.~Ohki,
  %"Three generation magnetized orbifold models,"
  Nucl.~Phys.~B {\bf 814}, 265 (2009)
% doi:10.1016/j.nuclphysb.2009.02.002
  [arXiv:0812.3534[hep-th]].
%%CITATION = doi:10.1016/j.nuclphysb.2009.02.002;%%

%\cite{Cremades:2004wa}
\bibitem{Cremades:2004wa} 
  D.~Cremades, L.~E.~Ibanez and F.~Marchesano,
  %``Computing Yukawa couplings from magnetized extra dimensions,''
  JHEP {\bf 0405}, 079 (2004)
% doi:10.1088/1126-6708/2004/05/079
[hep-th/0404229].
%%CITATION = doi:10.1088/1126-6708/2004/05/079;%%  


%\cite{Abe:2012fj}
\bibitem{Abe:2012fj} 
  H.~Abe, T.~Kobayashi, H.~Ohki, A.~Oikawa and K.~Sumita,
  %``Phenomenological aspects of 10D SYM theory with magnetized extra dimensions,''
  Nucl.\ Phys.\ B {\bf 870}, 30 (2013)
%  doi:10.1016/j.nuclphysb.2013.01.014
  [arXiv:1211.4317 [hep-ph]].
  %%CITATION = doi:10.1016/j.nuclphysb.2013.01.014;%%


  
%\cite{Abe:2014}
\bibitem{Abe:2014}
  H.~Abe, T.~Kobayashi, K.~Sumita and Y.~Tatsuta,
  %"Gaussian Froggatt-Nielsen mechanism on magnetized orbifolds,"
  Phys.~Rev.~D {\bf 90}, no.~10.~105006 (2014)
% doi:10.1103/PhysRevD.90.105006
  [arXiv:1405.5012[hep-ph]].
%%CITATION = doi:10.1103/PhysRevD.90.105006;%%

%\cite{Fujimoto:2016zjs}
\bibitem{Fujimoto:2016zjs} 
  Y.~Fujimoto, T.~Kobayashi, K.~Nishiwaki, M.~Sakamoto and Y.~Tatsuta,
  %``Comprehensive analysis of Yukawa hierarchies on $T^2/Z_N$ with magnetic fluxes,''
  Phys.\ Rev.\ D {\bf 94}, no. 3, 035031 (2016)
%  doi:10.1103/PhysRevD.94.035031
  [arXiv:1605.00140 [hep-ph]].
  %%CITATION = doi:10.1103/PhysRevD.94.035031;%%
  
 %\cite{Kobayashi:2016qag}
\bibitem{Kobayashi:2016qag} 
  T.~Kobayashi, K.~Nishiwaki and Y.~Tatsuta,
  %``CP-violating phase on magnetized toroidal orbifolds,''
  JHEP {\bf 1704}, 080 (2017)
%  doi:10.1007/JHEP04(2017)080
  [arXiv:1609.08608 [hep-th]].
  %%CITATION = doi:10.1007/JHEP04(2017)080;%% 

%\cite{Dixon:1985jw}
\bibitem{Dixon:1985jw}
  L.~J.~Dixon, J.~A.~Harvey, C.~Vafa and E.~Witten,
  %``Strings on Orbifolds,''
  Nucl.\ Phys.\ B {\bf 261} (1985) 678.
  
  %\cite{Dixon:1986jc}
\bibitem{Dixon:1986jc}
  L.~J.~Dixon, J.~A.~Harvey, C.~Vafa and E.~Witten,
  %``Strings on Orbifolds. 2.,''
  Nucl.\ Phys.\ B {\bf 274} (1986) 285.
  

%\cite{Nibbelink:2007rd}
\bibitem{Nibbelink:2007rd}
  S.~Groot Nibbelink, M.~Trapletti and M.~Walter,
  %``Resolutions of C**n/Z(n) Orbifolds, their U(1) Bundles, and Applications to String Model Building,''
  JHEP {\bf 0703} (2007) 035
  %doi:10.1088/1126-6708/2007/03/035
  [hep-th/0701227].  

%\cite{Leung:2019oln}
\bibitem{Leung:2019oln}
  P.~Leung and H.~Otsuka,
  %``Heterotic Stringy Corrections to Metrics of Toroidal Orbifolds and Their Resolutions,''
  Phys.\ Rev.\ D {\bf 99} (2019) no.12,  126011
  %doi:10.1103/PhysRevD.99.126011
  [arXiv:1903.12144 [hep-th]].


%\cite{Eguchi:1978xp}
\bibitem{Eguchi:1978xp}
  T.~Eguchi and A.~J.~Hanson,
  %``Asymptotically Flat Selfdual Solutions to Euclidean Gravity,''
  Phys.\ Lett.\  {\bf 74B} (1978) 249.


%\cite{Kobayashi:2019}
\bibitem{Kobayashi:2019}
  T.~Kobayashi, H.~Otsuka and H.~Uchida,
  %"Wavefunctions and Yukawa couplings on Resolutions of $T^2/Z_N$ Orbifolds,"
  JHEP {\bf 1908}, 046 (2019)
% doi:10.1007/JHEP08(2019)046
  [arXiv:1904.02867 [hep-th]].
%%CITATION = doi:10.1007/JHEP08(2019)046;%%







%\cite{Conlon:2008qi}
\bibitem{Conlon:2008qi}
  J.~P.~Conlon, A.~Maharana and F.~Quevedo,
  %``Wave Functions and Yukawa Couplings in Local String Compactifications,''
  JHEP {\bf 0809}, 104 (2009)
% doi:10.1088/1126-6708/2008/09/104
  [arXiv:0807.0789 [hep-th]].
%%CITATION = doi:10.1088/1126-6708/2008/09/104;%%
  


%\cite{Kobayashi:2015siy}
\bibitem{Kobayashi:2015siy} 
  T.~Kobayashi, Y.~Tatsuta and S.~Uemura,
  %``Majorana neutrino mass structure induced by rigid instantons on toroidal orbifold,''
  Phys.\ Rev.\ D {\bf 93}, no. 6, 065029 (2016)
%  doi:10.1103/PhysRevD.93.065029
  [arXiv:1511.09256 [hep-ph]].
  %%CITATION = doi:10.1103/PhysRevD.93.065029;%%




%\cite{Xing:2008}
\bibitem{Xing:2008}
  Z.~z.~Xing, H.~Zhang and S.~Zhou,
  %"Updated Values of Running Quark and Lepton Masses,"
  Phys.~Rev.~D {\bf 77}, 113016 (2008)
% doi:10.1103/PhysRevD.77.113016
  [arXiv:0712.1419[hep-ph]].
%%CITATION = doi:10.1103/PhysRevD.77.113016;%%  


%\cite{Tanabashi:2018oca}
\bibitem{Tanabashi:2018oca}
  M.~Tanabashi {\it et al.} [Particle Data Group],
  %``Review of Particle Physics,''
  Phys.\ Rev.\ D {\bf 98} (2018) no.3,  030001.
 % doi:10.1103/PhysRevD.98.030001

%\cite{Abe:2013bca}
\bibitem{Abe:2013bca} 
  T.~H.~Abe, Y.~Fujimoto, T.~Kobayashi, T.~Miura, K.~Nishiwaki and M.~Sakamoto,
  %``$Z_N$ twisted orbifold models with magnetic flux,''
  JHEP {\bf 1401}, 065 (2014)
%  doi:10.1007/JHEP01(2014)065
  [arXiv:1309.4925 [hep-th]].
  %%CITATION = doi:10.1007/JHEP01(2014)065;%%


%\cite{Abe:2014noa}
\bibitem{Abe:2014noa} 
  T.~h.~Abe, Y.~Fujimoto, T.~Kobayashi, T.~Miura, K.~Nishiwaki and M.~Sakamoto,
  %``Operator analysis of physical states on magnetized $T^{2}/Z_{N}$ orbifolds,''
  Nucl.\ Phys.\ B {\bf 890}, 442 (2014)
%  doi:10.1016/j.nuclphysb.2014.11.022
  [arXiv:1409.5421 [hep-th]].
  %%CITATION = doi:10.1016/j.nuclphysb.2014.11.022;%%


%\cite{Headrick:2005ch}
\bibitem{Headrick:2005ch}
  M.~Headrick and T.~Wiseman,
  %``Numerical Ricci-flat metrics on K3,''
  Class.\ Quant.\ Grav.\  {\bf 22} (2005) 4931
  %doi:10.1088/0264-9381/22/23/002
  [hep-th/0506129].
  
 

%\cite{Burgess:2008ri}
\bibitem{Burgess:2008ri}
  C.~P.~Burgess, J.~P.~Conlon, L.~Y.~Hung, C.~H.~Kom, A.~Maharana and F.~Quevedo,
  %``Continuous Global Symmetries and Hyperweak Interactions in String Compactifications,''
  JHEP {\bf 0807} (2008) 073
  %doi:10.1088/1126-6708/2008/07/073
  [arXiv:0805.4037 [hep-th]].
  

%\cite{Maharana:2011wx}
\bibitem{Maharana:2011wx}
  A.~Maharana,
  %``Symmetry Breaking Bulk Effects in Local D-brane Models,''
  JHEP {\bf 1206} (2012) 002
  %doi:10.1007/s13130-012-4129-0, 10.1007/JHEP06(2012)002
  [arXiv:1111.3047 [hep-th]].
    

\end{thebibliography}
\end{document}